\documentclass{aa}
\usepackage{graphicx}
\begin{document}

  \title{Stellar and Wind Properties of LMC WC4 stars}

   \titlerunning{Study of LMC WC4 stars}

   \subtitle{A metallicity dependence for Wolf-Rayet mass-loss rates\thanks
{Based on observations made with the
NASA-CNES-CSA {\it Far Ultraviolet Spectroscopic Explorer}, and NASA-ESA
{\it Hubble Space Telescope}. Also based on observations 
collected at the European Southern
Observatory in program 63.H-0683, and at the 
Australian National University Siding Spring Observatory.}}

   \author{Paul A. Crowther\inst{1}, Luc Dessart\inst{1, 2}, 
          D. John Hillier\inst{3}, Jay B. Abbott\inst{1}, 
\and Alex W. Fullerton\inst{4,5}}

  \authorrunning{Crowther et al.}

   \offprints{P. A. Crowther}

   \institute{Department of Physics \& Astronomy, UCL, Gower Street,
      London WC1E 6BT, UK\\      
     \email{pac@star.ucl.ac.uk, jba@star.ucl.ac.uk}
     \and Sterrenkundig Institut, Universiteit Utrecht, P.O. Box 80000,
              NL-3508 TA Utrecht, The Netherlands\\
     \email{L.Dessart@astro.uu.nl}
     \and Department of Physics \& Astronomy, University of Pittsburgh, 
             3941 O'Hara Street, PA 15260\\
     \email{jdh@phyast.pitt.edu}
     \and Dept. of Physics \& Astronomy, University of Victoria,
    P.O. Box 3055, Victoria, BC, V8W 3P6, Canada.
     \and    Department of Physics \& Astronomy, Johns Hopkins University, 
     3400 North Charles St., Baltimore, MD 21218\\
     \email{awf@pha.jhu.edu}}
   \date{Received ; accepted }

   \abstract{
We use ultraviolet space-based ({\it FUSE, HST}) and optical/IR 
ground-based 
(2.3m MSSSO, NTT) spectroscopy
to determine the physical parameters of six WC4-type Wolf-Rayet stars in the 
Large Magellanic Cloud. 
Stellar parameters are revised significantly relative to Gr\"{a}fener et al.
(1998)
based on improved observations and more sophisticated model atmosphere codes,
which account for line blanketing and clumping.
We find that stellar luminosities are revised upwards by up to 0.4 dex, with 
surface abundances spanning a lower range of 
0.1$\le$ C/He $\le$ 0.35 (20--45\% carbon by mass) and O/He$\le$0.06
($\leq$10\% oxygen by mass). \\
Relative to Galactic WC5--8 stars at known 
distance, and analysed in a similar manner, LMC WC4 stars 
possess systematically higher stellar luminosities, 
$\sim$0.2\,dex lower wind densities, yet a similar range 
of surface chemistries. We illustrate how the classification
C\,{\sc iii} $\lambda$5696 line
is extremely sensitive to wind density, such that this is
the principal difference 
between the subtype distribution of LMC and Galactic early-type WC 
stars.  Temperature differences do
play a role, but carbon abundance does not 
affect WC spectral types.
We illustrate the effect of varying temperature and mass-loss rate
on the WC spectral type for 
HD\,32257 (WC4, LMC) and HD\,156385 (WC7, Galaxy) which
possess similar abundances and luminosities.\\
Using the latest evolutionary models, pre-supernova
stellar masses in the range 11--19 $M_{\odot}$ are anticipated for 
LMC WC4 stars, with 7--14 
$M_{\odot}$ for Galactic WC stars with known distances. 
These values are consistent with pre-cursors of
bright Type-Ic supernovae such as SN\,1998bw (alias GRB 980425) for which
a minimum total mass of C and O of 14$M_{\odot}$ has been independently 
derived.
}

   \maketitle
%

\section{Introduction}

The ultimate fate of the most massive stars is likely to be a Type Ib
or Ic Supernova (SN) explosion. Type Ib's, conspicuous 
for the absence of 
hydrogen in their spectra, likely correspond to WN-type Wolf-Rayet 
(WR) stars, whilst WC or WO stars are thought to be responsible for 
Type Ic SN, since both hydrogen and helium are absent  from their (early) spectra.
Are  the properties of WC stars immediately prior to their proposed SN explosion 
consistent with that of such SN? Models of WR stars and SN are now 
sufficiently advanced to facilitate such comparisons for the first time.

Over the past couple of decades, spectroscopic tools for the quantitative
analysis of WR stars have advanced sufficiently (e.g. Hillier \& Miller
1998) to permit the reliable
determination of abundances (Herald et al. 2001), 
masses (De Marco et al. 2000) and ionizing 
fluxes (Crowther et al. 1999).  Armed with these tools, 
we are now in an  unprecedented position to investigate such stars in the
Local Group and beyond (e.g. Smartt et al. 2001).
Already, high quality observations spanning UV to IR wavelengths, 
are possible for individual stars in our Galaxy or the Magellanic Clouds
with new instruments, such as 
the {\it Far-Ultraviolet Spectroscopic Explorer} ({\it FUSE}, Moos et al. 
2000).

WR stars in the LMC have been the focus of several spectroscopic 
investigations. WN stars have been studied by Conti \& Massey (1989), whilst 
Hamann \& Koesterke (1998), and references therein, have 
determined their  quantitative properties, for which a negligible 
metallicity effect was remarked upon relative to Galactic counterparts (the
heavy metal content of the LMC is $\sim 0.4Z_{\odot}$, Dufour 1984). 
Torres et al. 
(1986), Smith et al. (1990) and Barzakos et al. (2001) have compared 
LMC WC stars, with Gr\"{a}fener et al. (1998) presenting studies of six 
WC4 stars. In the latter work, WC4 stars were found to possess remarkably
uniform (and high) carbon and oxygen abundances, plus 
a wide range of stellar luminosities. Results were inconclusive 
whether the mass-loss rates of LMC WC stars are similar to, or lower than,
Galactic counterparts. We return to the study of
these stars in this paper, since we possess improved and more extensive 
spectroscopy, including far-UV and near-IR datasets, plus
better modelling tools (Hillier \& Miller 1998).

\begin{table}
\caption{Observing log for 
LMC WC4 stars, including narrow-band photometry from 
Torres-Dodgen \& Massey (1988), and catalogue numbers from Breysacher
(1981, Br) and Breysacher et al. (1999, BAT).}                        
\label{table1}
\begin{tabular}{l@{\hspace{1mm}}r@{\hspace{1mm}}r@{\hspace{1mm}}c@{\hspace{1mm}}
r@{\hspace{1mm}}r@{\hspace{1mm}}r@{\hspace{1mm}}r@{\hspace{1mm}}r}\\
\hline
HD     &    Br  & BAT &  $v$   &  $b-v$ & {\it FUSE} & {\it HST} & MSSSO & 
NTT \\
\hline
32125  &  7   &  9    & 15.02  & 0.10    & --  & Sep 94 &Dec 97 & -- \\
32257  &  8   &  8    & 14.89  & 0.13    & --  & Nov 94 &Dec 97 & -- \\
32402  & 10   &  11   & 13.89  & $-$0.06 &Nov 01  & Apr 95 &Dec 97 & Sep 99\\
37026  & 43   &  52   & 14.04  & $-$0.03 &  Feb 00  & Nov 94 &Dec 97 & -- \\
37680  & 50   &  61   & 14.03  & $-$0.01 &  Feb 00 & Nov 94 &Dec 97 & -- \\
269888 & 74   & 90    & 15.41  & 0.19    & --  & Jun 95 &Dec 97 & -- \\
\hline
\end{tabular}
\end{table}

We present our new observations in 
Sect.~2, and discuss the present set of model calculations in Sect.~3.
Individual results are presented in Sect.~4 and compared with the
previous study of Gr\"{a}fener et al. Quantitative comparisons are
made with
Galactic counterparts that have been 
analysed in a similar manner (Dessart et al. 2000;
Hillier \& Miller 1999) in Sect.~5, together with evolutionary
expectations. Finally, the 
possibility that WC stars provide the precursors to Type Ic SN are 
discussed in Sect.~6, via the comparison of accurate WC masses
with those of the CO-cores determined for recent SN Ic explosions 
(e.g. Iwamoto et al. 1998, 2000).

\begin{figure*}[ht!] 
\vspace{10cm} 
\includegraphics{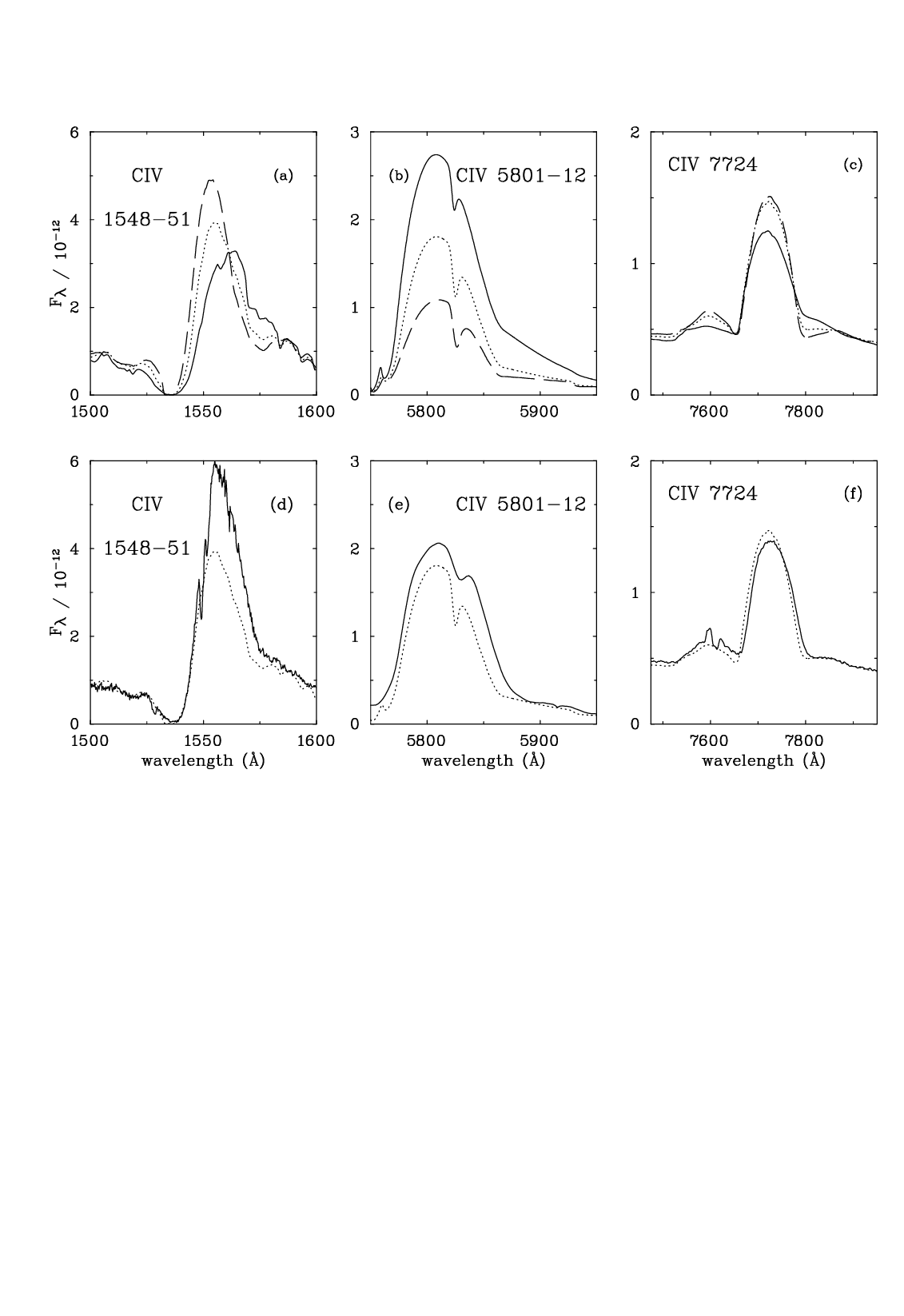}
\caption{{\it Panels (a-c)}: Synthetic spectra for
selected C\,{\sc iv} line profiles with parameters $T_{\star}=90$kK, $\log
L/L_{\odot}$=5.70, $v_{\infty}$=3000 km\,s$^{-1}$, C/He=0.14, O/He=0.02 
by number. An
unclumped model is shown in solid ($f$=1, with $\dot{M} = 1.1 \times 10^{-4}
M_{\odot}$ yr$^{-1})$, moderately clumped as a dotted line ($f$=0.1, with
$\dot{M} = 3.5 \times 10^{-5} M_{\odot}$ yr$^{-1})$), and highly clumped as a
dashed-line ($f=0.01$, with $\dot{M} = 1.1 \times 10^{-5} M_{\odot}$ yr$^{-1}$.
Electron scattering wings, and line intensities for the case of $\lambda\lambda$5801--12,
are very sensitive to clumping, as is the shape and location of
$\lambda\lambda$1548--51. {\it Panels (d-f):} Comparison of moderately clumped ($f$=0.1) 
models with observations of HD\,32402, showing generally good agreement. 
Ordinate units are erg\,cm$^{-2}$\,s$^{-1}$\,\AA$^{-1}$.}
\label{clump}
\end{figure*}

    \section{Observations}\label{sect2}

For our study, we utilise archival UV {\it Hubble Space Telescope
(HST)} datasets, together with previously unpublished far-UV {\it FUSE}, 
optical
2.3m Mt Stromlo \& Siding Spring Observatory (MSSSO) spectroscopy, plus near-IR
European Southern Observatory (ESO) 
New Technology Telescope (NTT) observations, discussed in turn below. 
An observing log for each of the six program stars, 
together with photometry from 
Torres-Dodgen \& Massey (1988), is presented in Table~\ref{table1}.
Spectral types are WC4 in all cases, following either Smith et al. (1990)
or Crowther et al. (1998).

\subsection{Far UV spectroscopy}

HD 32402, HD 37026, and HD 37680 were observed as part of the
{\it FUSE} Principal Investigator team program to study hot stars (P117, 
P.I.: J. B. Hutchings) between 2000 February--2001 November, as indicated
in Table~\ref{table1}.
The spectra were obtained as time-tag mode integrations of
4.5--8.1 ks duration through the $30\arcsec \times 30\arcsec$ (LWRS)
aperture.
As described by Moos et al. (2000) and Sahnow et al. (2000), {\it FUSE}
data consist of spectra with spectral resolution of $\sim$15,000 from
two Lithium Fluoride (LiF) channels, which cover
$\lambda\lambda$ 990--1187~{\AA} and two Silicon Carbide (SiC) channels,
which cover $\lambda\lambda$ 905--1105~{\AA}.
Spectra from each channel were processed by the current version of the
standard calibration pipeline ({\sc calfuse} 2.0.5), which corrects for 
drifts
and distortions in the readout electronics of the detectors, removes the
effects of thermally induced grating motions, subtracts a background image,
corrects for residual astigmatism in the spectrograph optics, and applies
flux and wavelength calibrations to the extracted spectra.
Spectra from the individual channels were subsequently aligned, merged, and
resampled to a constant wavelength step of 0.1~{\AA} in the manner
described by Walborn et al. (2002b).

\subsection{Near UV spectroscopy}

All program stars were observed with the {\it HST} Faint Object 
Spectrograph 
(FOS) instrument between 1994 September--1995 June (PI: D.J. Hillier). 
Individual exposures were obtained with
the G130H, G190H and G270H 
gratings, respectively, providing complete spectral coverage in the
$\lambda\lambda$1140--3301 range. These data have previously been discussed
and published in part by Gr\"{a}fener et al. (1998).

\subsection{Optical and far-red spectroscopy}

We have used the Double Beam Spectrograph (DBS) at the 2.3m MSSSO
telescope  to observe our six targets between 1997 Dec 24--27. 
Nearly complete coverage in the optical and far-red region was obtained in
two separate exposures, which were in the range 240-1500 sec. The dichroic
together 
with  the 300l/mm (blue) and 316l/mm (red) gratings 
provided spectroscopy of 3620--6085\AA\ and 6410--8770\AA,
whilst the 600l/mm (blue) and 316 l/mm (red) 
gratings permitted observations at 
3240--4480\AA\  and 8640--11010\AA\  simultaneously. 
The detectors for both arms of the DBS were 
1752$\times$532 pixel SITE CCD's. A
2$''$ slit provided a
2 pixel spectral resolution of $\sim$5\AA, whilst a 6$''$ slit (for
absolute spectrophotometry) provided a spectral resolution of $\sim$12\AA.
A standard data reduction was  carried out, including absolute 
flux calibration, using wide slit spectrophotometry of 
HD\,60753 (B3\,IV) and $\mu$ Col (O9\,V), plus atmospheric correction 
using B stars HR\,2221, 4074 and 4942.

\subsection{Near-IR spectroscopy}

Long slit, near-IR spectroscopy of HD\,32402 was acquired on  1999
Sept 1--2 with the NTT, using the Son of Isaac (SofI) instrument, 
a 1024$\times$1024 pixel NICMOS detector, and low resolution IJ (GRB) 
grating, with spectral coverage 0.94--1.65~$\mu$m and
a dispersion of 7.0~\AA/pix. 
The 0.6$''$ slit provided a 2-pixel spectral resolution
of 14\AA. The total integration time was 480 sec (1 Sept) and 960 sec
(2 Sept).  
The removal of atmospheric features was achieved by observing
HD~38150 (F8\,V)
immediately after HD\,32402, at a close airmass (within 0.03).
Similar observations of Hip 22663 (B2\,V) permitted a relative flux
correction, using a $T_{\rm eff}$=22kK Kurucz model atmosphere normalized 
to V=7.66\,mag.

A standard extraction and wavelength calibration was carried out with
{\sc iraf}, while {\sc figaro} (Shortridge, Meyerdierks \& Currie, 1999)
and {\sc dipso} (Howarth et al. 1998) were used for the  atmospheric
and flux calibration, first artificially removing stellar hydrogen features
from the HD\,38150 spectrum. Our relatively fluxed dataset was 
adjusted to match MSSSO spectrophotometry.

\begin{table*}
\begin{center}
\caption[]{Summary of the WC model atom calculations.
N$_{\sc f}$ is the number of full levels, N$_{\sc s}$ the number 
of super levels and N$_{\sc t}$ the corresponding number of 
transitions. A total of 2526 levels (788 super levels) and 26\,239 transitions
are considered.}
\label{tabatom}
\begin{tabular}{
l@{\hspace{1mm}}
c@{\hspace{1mm}}c@{\hspace{1mm}}c
c@{\hspace{1mm}}c@{\hspace{1mm}}c
c@{\hspace{1mm}}c@{\hspace{1mm}}c
c@{\hspace{1mm}}c@{\hspace{1mm}}c
c@{\hspace{1mm}}c@{\hspace{1mm}}c
c@{\hspace{1mm}}c@{\hspace{1mm}}c
c@{\hspace{1mm}}c@{\hspace{1mm}}c
c@{\hspace{1mm}}c@{\hspace{1mm}}c}
\hline
& \multicolumn{3}{c}{I} & 
\multicolumn{3}{c}{II} & 
\multicolumn{3}{c}{III} & 
\multicolumn{3}{c}{IV} & 
\multicolumn{3}{c}{V} & 
\multicolumn{3}{c}{VI} & 
\multicolumn{3}{c}{VII} &
\multicolumn{3}{c}{VIII} \\

Ion          &      
$N_{\sc f}$  &  $N_{\sc s}$ & $N_{\sc t}$ &
$N_{\sc f}$  &  $N_{\sc s}$ & $N_{\sc t}$ &
$N_{\sc f}$  &  $N_{\sc s}$ & $N_{\sc t}$ &
$N_{\sc f}$  &  $N_{\sc s}$ & $N_{\sc t}$ &
$N_{\sc f}$  &  $N_{\sc s}$ & $N_{\sc t}$ &
$N_{\sc f}$  &  $N_{\sc s}$ & $N_{\sc t}$ &
$N_{\sc f}$  &  $N_{\sc s}$ & $N_{\sc t}$ &
$N_{\sc f}$  &  $N_{\sc s}$ & $N_{\sc t}$ \\
\hline 
 He & 39 & 27 &315 & 30 & 13 &435 \\
  C &    &    &    & 16 &  9 & 36  & 243 & 100& 5513& 64 & 49 &1446\\
  O &    &    &    & 29 & 13 & 120& 154 & 60 & 1454& 72 & 30 &835 & 78 & 41 &524 &23 & 17 & 109\\
 Ne &    &    &    & 48 & 14 & 290& 101 & 39 & 675 & 68 & 23 &432 & 117& 27 &837 &34 & 15 &135 \\
 Si &    &    &    &    &    &    &     &    &     & 33 & 23 &183\\
 P  &    &    &    &    &    &    &     &    &     & 28 & 16 & 57 & 28 & 18 &138  \\
 S  &    &    &    &    &    &    &     &    &     & 77 & 29 &506 & 26 & 14 & 58 &11 &  7 & 24  \\
 Ar &    &    &    &    &    &    & 36  & 10 & 67  & 105& 23 & 834& 61 & 24 & 276&81 & 21 &606& 37& 18 & 133& 12& 7 &26 \\
 Fe &    &    &    &    &    &    &     &    &     &280 & 21 &4223&182 & 19 &2163&190& 29 &2028&153& 14&1216& 70& 18&483\\
\hline
\end{tabular}
\end{center}
\end{table*}

\section{Model Analysis}\label{sect3}

\subsection{Method}

For this study, we employed the code of Hillier \& Miller (1998), 
{\sc cmfgen}, which 
iteratively solves the  transfer equation in the co-moving 
frame subject to statistical and radiative  equilibria in an expanding, 
spherically symmetric and steady-state atmosphere. These models account
for clumping, via a volume filling factor, $f$,  and line blanketing,
since these effects affect significantly the derived 
properties of WC stars (e.g. Hillier \& Miller 1999). Through the 
use of `super-levels', extremely  complex atoms can been included. 
For the present application, a total 
of 2526 levels (combined into 788 super-levels), 40 depth points and 
26\,239 spectral lines of He, C, O, Ne, Si, S, P, Ar and Fe are considered
as indicated in Table~\ref{tabatom}.  The reader is referred to 
Dessart et al. (2000) for details of the oscillator strengths, collision
and photoionization cross-sections. The OPACITY (Seaton 1987, 1995) and
IRON projects (Hummer et al. 1993) 
formed the basis for most atomic data utilised in this work. 
Test calculations,
additionally involving extensive model atoms for 
Na\,{\sc ii-vii}, 
 Mg\,{\sc iii-vii}, Cl\,{\sc iv-vii}, Ca\,{\sc iv-ix}, Cr\,{\sc iv-vi}, Mn\,{\sc iv-vii}
and Ni\,{\sc iv-ix} were also considered, with minor effects on the emergent
spectrum, except for an increase in the strength of C\,{\sc iv} $\lambda\lambda$1548--51, 5801--12.

Following theoretical discussions by Schmutz (1997), and for consistency with
previous studies, we adopt a form for the  
velocity law (Eqn~8 from Hillier \& Miller 1999) such that two exponents 
are considered ($\beta_{1}$=1, $\beta_{2}$=20), together with an 
intermediate ($v_{\rm ext}$) and terminal $v_{\infty}$ velocity, with
$v_{\rm ext} \sim v_{\infty}$ - 300 km\,s$^{-1}$. Consequently,
acceleration is modest at small radii, and continues to relatively 
large distances, i.e. 0.9$v_{\infty}$ is reached at $\sim$20$R_{\ast}$
versus 10$R_{\ast}$ for a standard $\beta$=1 law. 
In the future, we intend to relax this assumption via hydrodynamical
driving, at least for the outer envelope. Terminal 
velocities are measured principally from the C\,{\sc iv}
$\lambda\lambda$1548--51 UV P Cygni profiles, subject to minor adjustment at the
line fitting stage, and agree relatively well (within 15\%) 
with previous determinations from Gr\"{a}fener et al. (1998) who
measured wind velocities using the unsaturated C\,{\sc iii} $\lambda$2297 line.

The formal solution of the radiative transfer equation, yielding the final
emergent spectrum, is computed separately. Generally, Doppler profiles are
adopted throughout, 
except for C\,{\sc iii} $\lambda$977, C\,{\sc iv} $\lambda\lambda$1548--51
for which  Voigt profiles are absolutely essential. 
Except where noted,
these final calculations assume $v_{\rm turb}$=50 km\,s$^{-1}$.

\begin{figure}[ht!]
\vspace{14cm}
\includegraphics{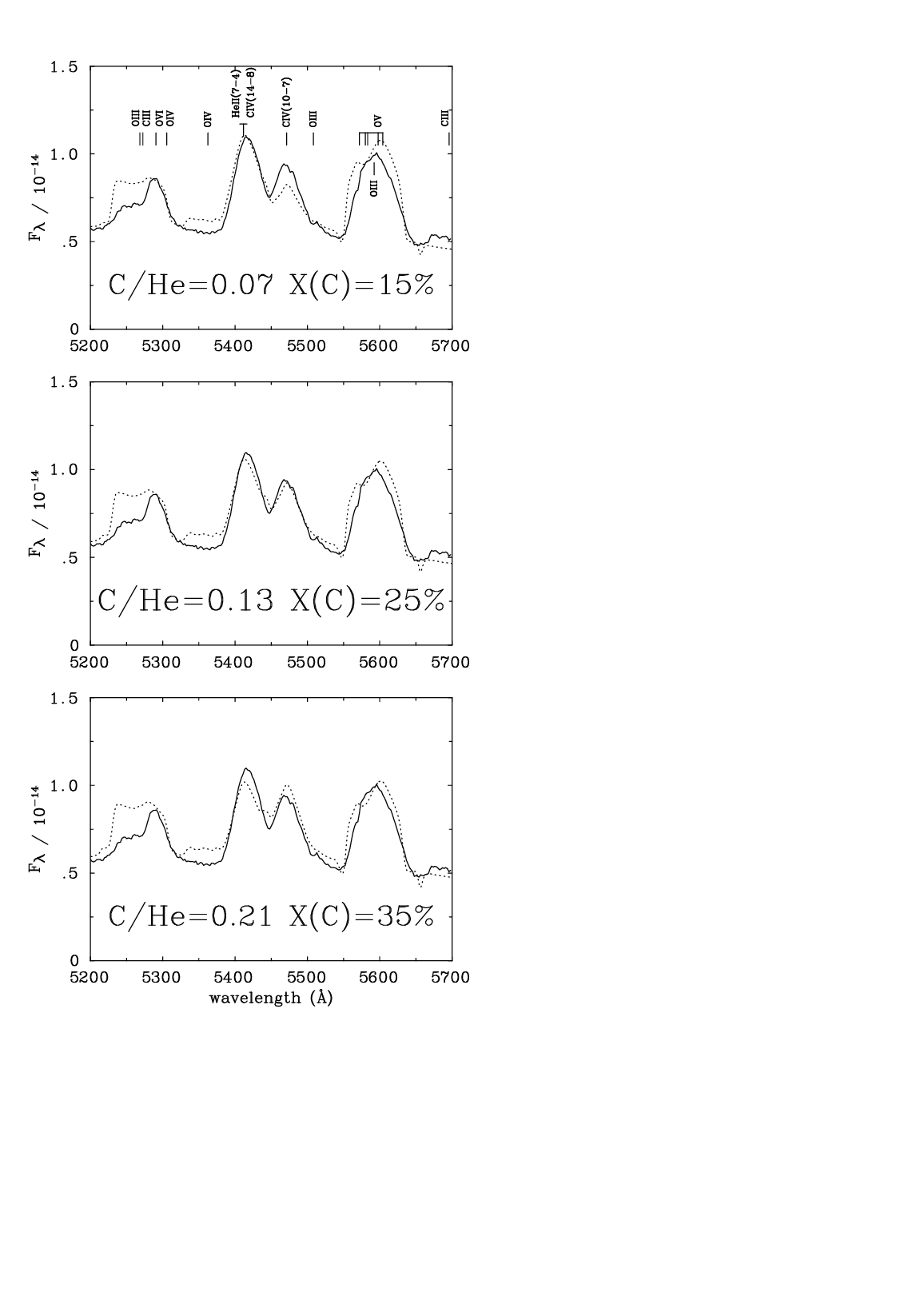}
\caption{Comparison between the de-reddened spectroscopy of 
HD\,32125 (solid) in the He\,{\sc ii} 5412 and C\,{\sc iv} 5471 
region, and theoretical predictions (dotted). In addition to those
features identified ($W_{\lambda}>$5\AA), there are numerous 
weaker carbon, oxygen, neon and sulphur features, From 
top-to-bottom:
C/He=0.07 by number (15\% by mass), 0.13 (25\%, adopted) and 0.21 (35\%). 
Ordinate units are erg\,cm$^{-2}$\,s$^{-1}$\,\AA$^{-1}$. }
\label{C_He}
\end{figure}

A series of models were calculated for each star in which stellar
parameters, $T_{\ast}$, log\,$(L/L_{\odot})$,
$\dot{M}/\sqrt{f}$, C/He and O/He, were adjusted until  the
observed line strengths and spectroscopic fluxes were
reproduced. Our spectroscopic analysis derives $\dot{M}/\sqrt{f}$,
rather than $\dot{M}$ and $f$ individually, since
line blending is so severe in WC4 stars. This is illustrated in
Fig.~\ref{clump}(a-c), where we compare selected synthetic line profiles
using identical models of HD\,32402, 
except that volume filling factors of $f$=1.0, 0.1 and 0.01 are adopted (with
corresponding changes in $\dot{M}$). C\,{\sc iv} $\lambda$7724 is typical
of the vast majority of lines, in that solely changes in electron 
scattering wings are predicted. Emission intensities of some 
lines, such as $\lambda\lambda$5801--12, are very sensitive to clumping, 
as are the shapes of the carbon resonance lines C\,{\sc iv} 
$\lambda\lambda$1548--51 and C\,{\sc iii} $\lambda$977. 
Recall that a Voigt profile is necessary for these lines. It
is apparent from Fig.~\ref{clump}(d-f) that the $f$=0.1 case
provides the best overall match to observed electron scattering 
wings, and to the central wavelength of 
$\lambda\lambda$1548--51, which unusually distinguishes itself as a 
clumping probe in WC stars in this way.

Hamann \& Koesterke (1998) considered clumped models of
HD\,37026, with
$f$=1.0, 0.25 and 0.0625, and came to the conclusion that the latter provided
the best fit to observations. In general, we can firmly
rule out $f$=1.0 for all cases, but our (admittedly simplistic) treatment
of wind clumping prevents a precise determination -- see Dessart \& 
Owocki (2002ab) for recent progress in this field.
Ultimately, we cannot pretend to derive the clumpiness of the wind, and
so we have selected identical values of $f$(=0.1) for the other program stars, 
as a reasonable compromise to the observed spectra in each case, and for
consistency with other (Galactic) stars analysed in our previous studies.

\subsection{Stellar Parameters and Abundances} 

The wind ionization balance is ideally deduced using isolated spectral 
lines from adjacent ionization stages  of helium, carbon or oxygen.
In order to ensure consistency with studies that are restricted to
optical or far-red wavelengths we selected He\,{\sc i} 10830/He\,{\sc ii} 
5412 and C\,{\sc iii} 6740+8500+9710/C\,{\sc iv} 7700. 
No suitable oxygen diagnostics were available in the optical or far-red, 
whilst use of He\,{\sc i} 10830 
was difficult in practise for our program stars 
(given the poor CCD response in the near-IR), with the exception of 
NTT-SofI data for  HD\,32402. 
Consequently, carbon provided
our principal diagnostic lines, with the carbon-to-helium abundance
determined from C\,{\sc iv} 5471 and He\,{\sc ii} 5412. We consider
the accuracy of our derived luminosities to be 15\%, with effective
temperatures somewhat worse constrained. Mass-loss rates also have a
15\% accuracy if the adopted filling factor if correct.

\begin{figure}[ht!]
\vspace{13cm}
\includegraphics{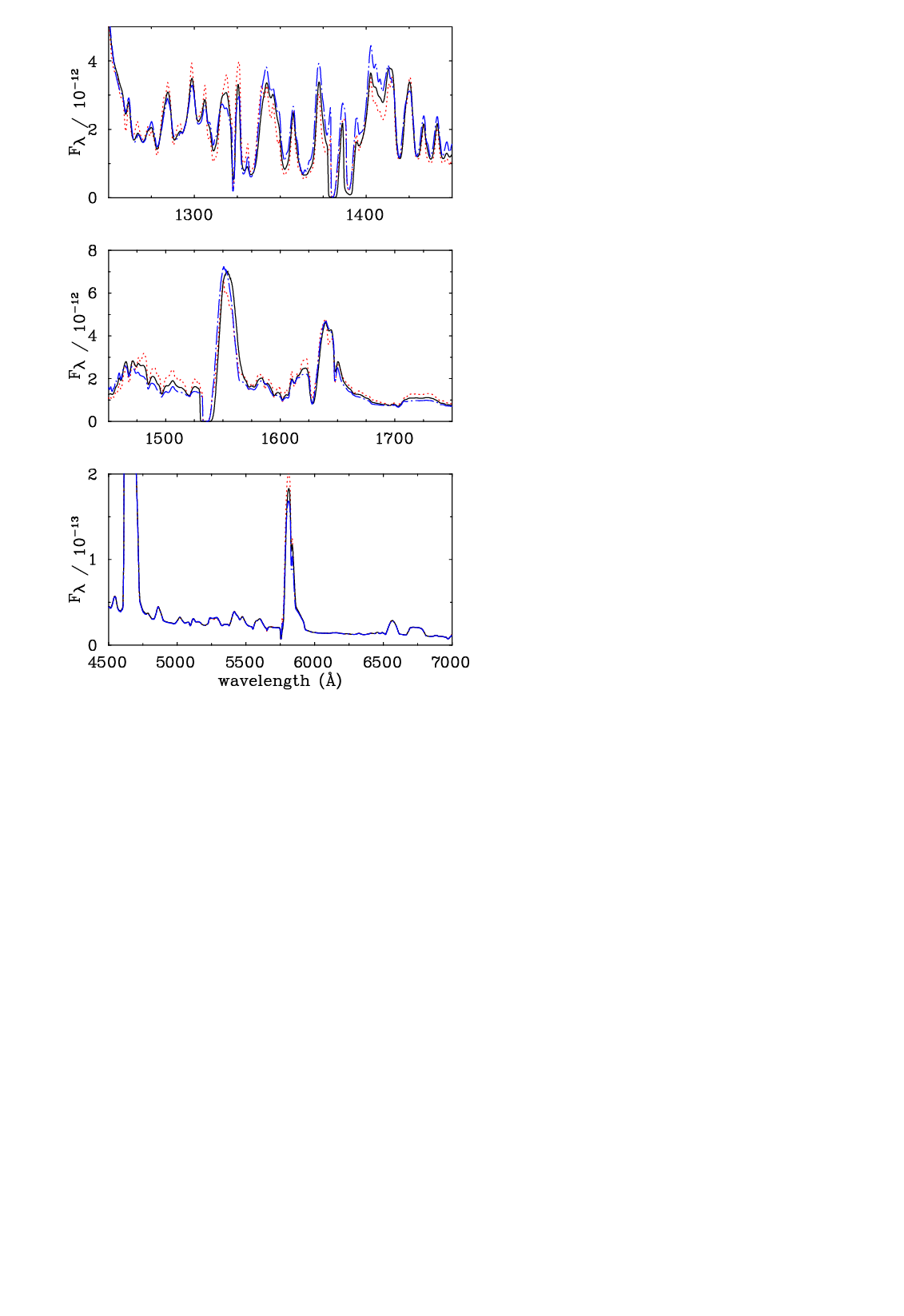}
\caption{Comparison between identical synthetic spectra for HD\,32402, except 
that iron abundances are chosen to be 0.2$Z_{\odot}$ (dashed, blue), 
0.4$Z_{\odot}$ (solid, black) and 0.8$Z_{\odot}$ (dotted, red). This illustrates how difficult it is
to accurately determine iron abundances based on UV spectroscopy in WC stars. 
Differences in iron content within this range has little effect
on the emergent carbon-oxygen-helium (diagnostic) lines.
Ordinate units are erg\,cm$^{-2}$\,s$^{-1}$\,\AA$^{-1}$.}
\label{metal}
\end{figure}

In Fig.~\ref{C_He} we present de-reddened observations
of HD\,32125, plus three identical models except that 
C/He ratios are 0.07, 0.13 (adopted) and 0.21 by number, 
corresponding to 15\%, 25\% and 
35\% carbon by mass, with the oxygen mass 
fraction maintained at 10\%. The relatively
small changes in line strength for differing C/He ratios emphasise
the need for high S/N and absolute flux calibration. 
Difficulties are
only encountered for stars with very large wind velocities in which
C\,{\sc iv} 5471 and He\,{\sc ii} 5412 are severely blended.
Overall, we consider an (internal) accuracy of 10\% in our derived
carbon mass fraction, i.e. 25$\pm$3\% for the case illustrated above.
External errors are difficult to quantify, but from test calculations
carried out using the line blanketed code of Gr\"afener et al. (2002)
the degree of consistency is excellent.
 
\begin{figure*}[t!] 
\vspace{21.75cm}
\includegraphics{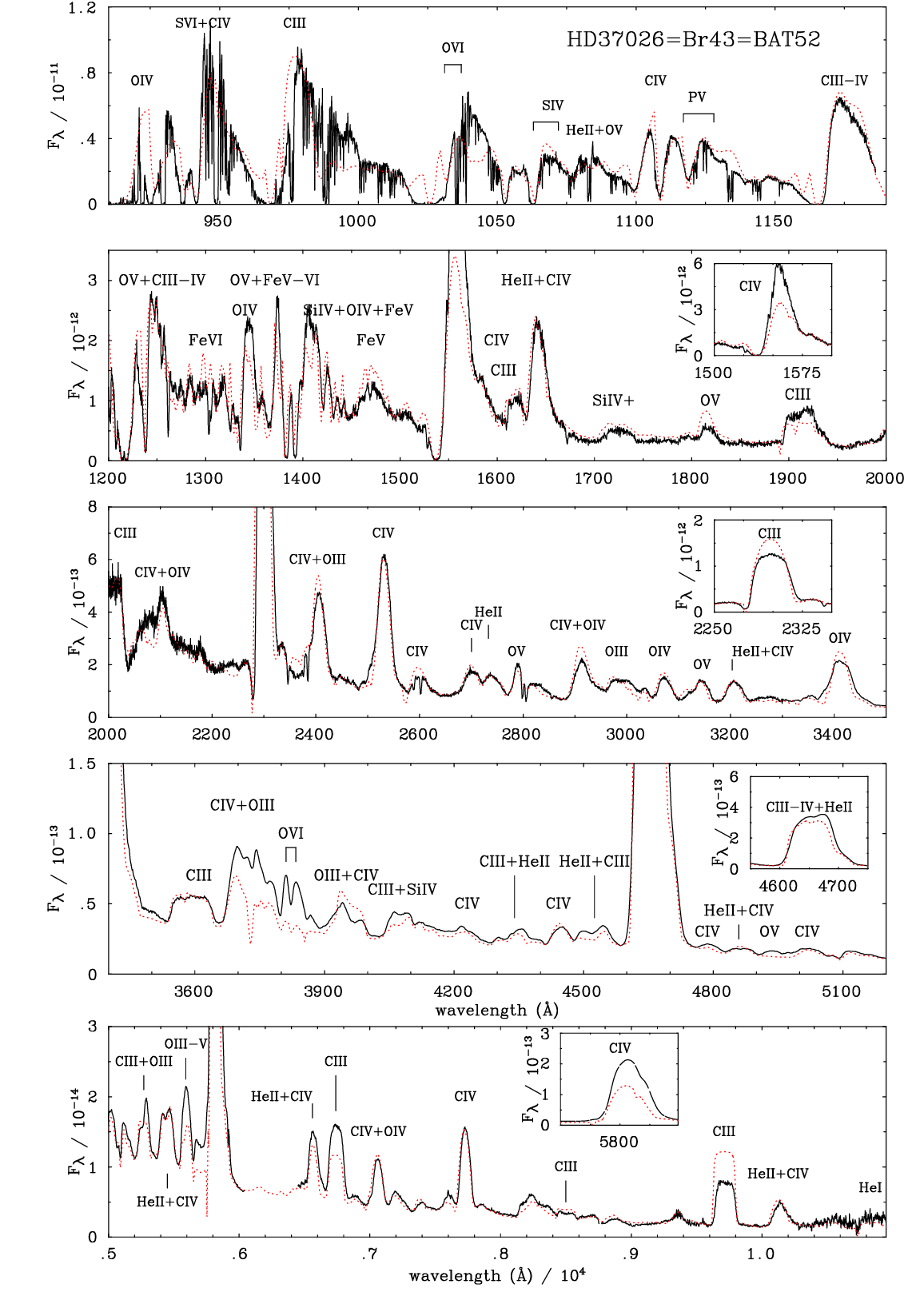}
\caption{Comparison between the de-reddened 
($E_{\rm B-V}$=0.07(Galactic)+0.04(LMC)) flux distribution of HD\,37026 
(solid, black) and theoretical predictions (dotted, red)
for $T_{\star}=90$kK, $\log L/L_{\odot}$=5.65,
$\log \dot{M}/(M_{\odot}$ yr$^{-1}) = -4.5$, $v_{\infty}$=2900 km\,s$^{-1}$, 
C/He=0.32, O/He=0.05 by number. For the {\it FUSE} region, we use an 
atomic 
hydrogen column density of $\log N$(H/cm$^{-2}$) = 21.0, whilst correction for 
molecular hydrogen is taken from $H_{2}$ models by Tumlinson et al. (2002) using
a Doppler width ($b=5.4$ km\,s$^{-1}$) and total column of
log (H$_{2}$/cm$^{-2}$)=15.45, considering levels $J$=0 to 3.
Ordinate units are erg\,cm$^{-2}$\,s$^{-1}$\,\AA$^{-1}$, whilst abscissa
units are \AA\, except for $\mu$m in the bottom panel.}
\label{br43_fit}
\end{figure*}

In contrast with carbon, the determination of oxygen abundances are
much more difficult. This is due to the absence of 
recombination lines amenable to simple analysis, a 
much more complex ionization balance structure than He and C, and more
problematic line blending -- for example the well known oxygen feature
between C\,{\sc iv} $\lambda$5471 and C\,{\sc iii} $\lambda$5696
is almost equally split between O\,{\sc iii} $\lambda$5592 and O\,{\sc v}
$\lambda\lambda$5572-5607. Nevertheless, there are numerous UV and optical
lines of oxygen which are sensitive to abundance variations, which
are discussed below for HD\,37026.
For oxygen we admit an (again internal) accuracy of 50\% in our 
derived oxygen mass fractions. Absolutely, a factor of two is probably
a more objective value. Again, test calculations by
Gr\"afener (priv. comm.) indicates good agreement, except for O\,{\sc iii}
lines for which dielectronic contributions to lines differ somewhat.

For elemental species other than He, C and O, 
we generally adopt 0.4$Z_{\odot}$ 
abundances (Si, S, P, Ar and Fe). 
We have considered models in which the iron
abundance is doubled or halved, without particularly affecting the emergent
spectrum.  This is illustrated in Fig.~\ref{metal} for UV synthetic spectra of
HD\,32402. For the case of neon, there is observational evidence that Ne is
enriched in Galactic WC stars such that its mass fraction is $\sim$1.5\%
(Ne/He=0.004 by number, Dessart et al. 2000). Consequently, we have sought
UV/optical line diagnostics as evidence for Ne enrichment in LMC WC4 stars
(see below).

\subsection{Reddening and Distance}

Simultaneously with the determination of the
ionization balance, wind density
and chemistry of each star, the luminosity is adjusted to ensure consistency
with de-reddened spectroscopy. In all cases, a Galactic foreground extinction
of $E_{\rm B-V}$=0.07 is adopted following Schlegel et al. (1998), using the
parameterization of Seaton (1979), with  $R$=3.1=$A_{\rm V}/E_{\rm B-V}$.
The extinction law followed by Seaton differs substantially from
Cardelli et al. (1989) for $\lambda\leq$1500\AA.
  The distance modulus to the LMC was assumed to be 18.50, as derived by
Cepheids and SN 1987A (Lee et al. 1993; Panagia et al. 1991), corresponding to 
50\,kpc.

LMC contributions are varied in $E_{\rm B-V}$ and $R$ following 
the parameterization of Howarth (1983). In the far-UV, interstellar
absorption in the H\,{\sc i} atomic Lyman series is accounted for 
following the technique of
Herald et al. (2001). Correction for molecular hydrogen in the 
{\it FUSE} spectral region has been made by using H$_{2}$ column densities
determined by Tumlinson et al. (2002): log (H$_{2}$/cm$^{-2}$)=15.45 and 
18.94 for HD\,37026 and HD\,37680, respectively. The H$_{2}$ column
density towards HD\,32402 is substantial, with
log (H$_{2}$/cm$^{-2}$) $\sim$19. HD\,37026, therefore, has by far the
`cleanest' far-UV stellar spectrum of the three WC4 stars thus far
observed with FUSE.

   \section{Results for individual stars} \label{sect4}

We have decided against presenting spectral fits for each star in turn,
since the model successes and failures are generally common to all 
our program stars. We solely present spectral fits to the entire 
far-UV, UV,  optical and near-IR observations of one star, HD\,37026 
in Fig.~\ref{br43_fit}. 

\subsection{Study of HD\,37026 (Brey~43 = BAT99-52)}

\begin{figure}[ht!]
\vspace{8.5cm}
\includegraphics{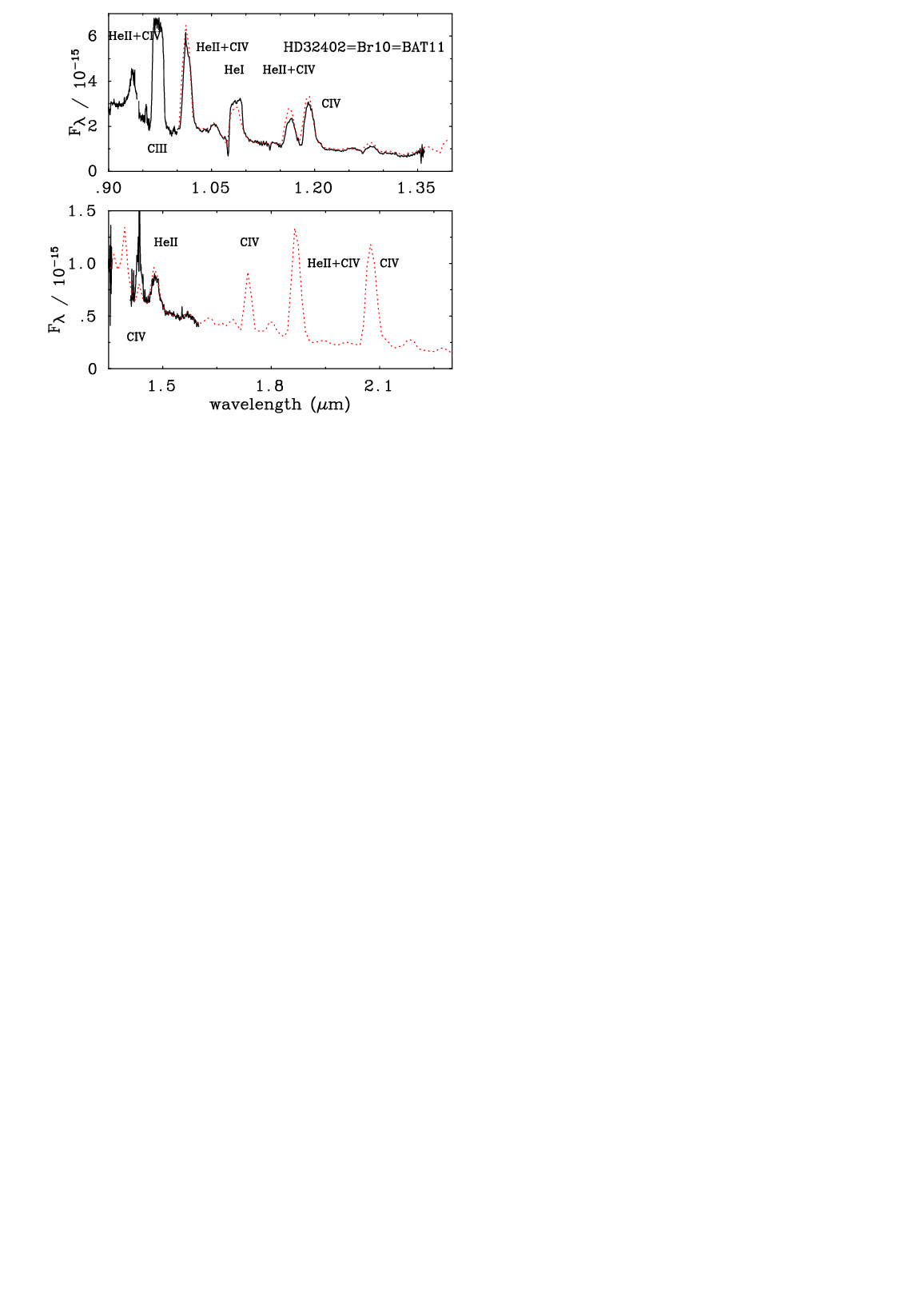}
\caption{Comparison between de-reddened near-IR
NTT-SofI observations of HD\,32402 (solid, black) and
synthetic spectra (dotted, red) determined from optical diagnostics.
Ordinate units are erg\,cm$^{-2}$\,s$^{-1}$\,\AA$^{-1}$.}
\label{br10ir}
\end{figure}

Overall, the 
flux distribution of HD\,37026
is very well reproduced by the synthetic model -- 
particularly the iron forest 
in the $\lambda\lambda$1250--1500 region and plethora of near-UV and
optical carbon-helium features with C/He=0.32 by number (44\% by mass).
He\,{\sc i} $\lambda$10830 may be rather underestimated in strength, 
although the low S/N of the observations, due
to the poor CCD response beyond 1$\mu$m, hinders detailed comparisons.
To illustrate the reliability of near-IR synthetic spectra for WC4
stars based on studies from optical diagnostics we present high S/N
NTT-SofI observations of HD\,32402 in Fig.~\ref{br10ir} with model
predictions. Overall the match is excellent, including He\,{\sc i} 
$\lambda$10830.

\begin{figure*}[t!]
\vspace{12cm}
\includegraphics{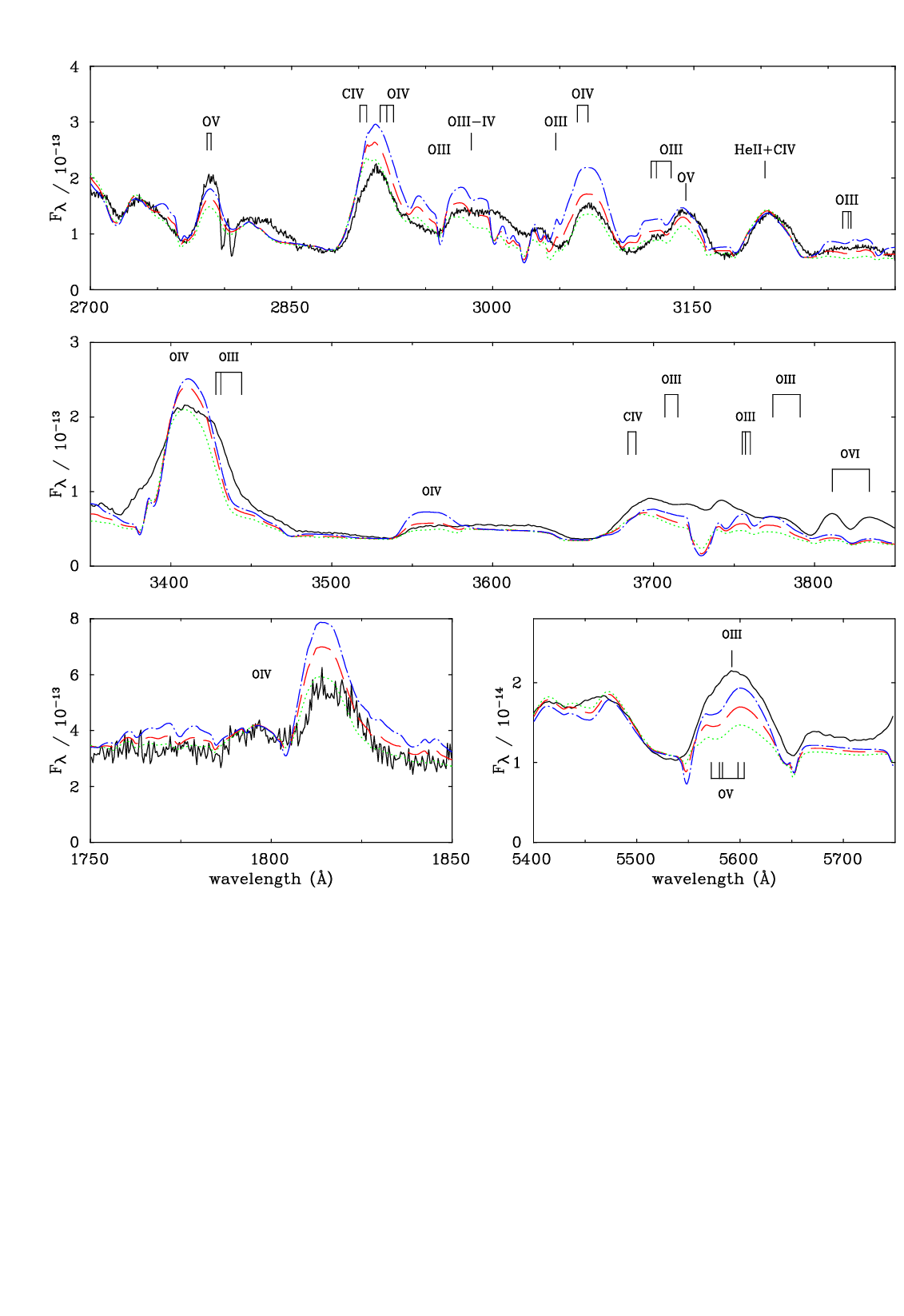}
\caption{Comparison between de-reddened observations of HD\,37026 (solid, black)
and synthetic spectra for a variety of oxygen abundances -- 5\% by mass
(dotted, green), 10\% (dashed, red) and 20\% (dot-dash, blue). O\,{\sc iv} $\lambda\lambda$3403--13
is rather insensitive to abundance, whilst the strong $\lambda$2900 feature
is a blend of C\,{\sc iv} $\lambda$2906, 2918 and O\,{\sc iv} $\lambda$2921.
Ordinate units are erg\,cm$^{-2}$\,s$^{-1}$\,\AA$^{-1}$.}
\label{br43-oxy}
\end{figure*}

Problematic carbon features for
HD\,37026 presented in Fig.~\ref{br43_fit} are as follows. 
C\,{\sc iv} $\lambda\lambda$1548--51
emission is  predicted  to be somewhat too weak. 
Previous difficulties with reproducing the
C\,{\sc iv} $\lambda\lambda$5801-12 feature (e.g. Hillier \& Miller 1999)
appear to have been mostly resolved by
the inclusion of additional metal opacity (specifically Ne and Ar).
Unfortunately, C\,{\sc iii} lines provide conflicting information.
For example, $\lambda$977, $\lambda$1909, $\lambda$2297,
$\lambda\lambda$4647--50 and $\lambda$8500 are reproduced to within 25\%
in HD\,37026. However,  $\lambda$6470 is up to 50\% too weak, 
whilst  $\lambda$9710 is similarly too strong. The blend including 
C\,{\sc iii} $\lambda$3747 is systematically too weak, despite 
efforts to improve atomic data. Amongst other WC4 stars, consistency between
$\lambda$6470 and  $\lambda$9710 is generally excellent.

As introduced above, there are numerous UV and optical
lines of oxygen which are sensitive to abundances variations although
results are somewhat discrepant.
Fig.~\ref{br43-oxy} presents identical models for HD\,37026 
except that the oxygen mass fraction is 5\% (O/He=0.025), 10\% (O/He=0.05) or
20\% (O/He=0.1). Diagnostics which correlate well with abundance
are O\,{\sc v} $\lambda$1815 (indicating 5\%), O\,{\sc iii} 
$\lambda$2983 (10\%),
O\,{\sc iv} $\lambda$3072 ($\sim$8\%), O\,{\sc iv} $\lambda\lambda$3560-3 ($\sim$8\%)
and O\,{\sc iii+v} $\lambda$5590 ($>$20\%). It is unfortunate that for
HD\,37026, the only strong oxygen feature available in the visual implies
a substantially different oxygen abundance from the numerous other diagnostics.
This is not always the case -- $\lambda$5590 is representative of near-UV
diagnostics for HD\,32125, HD\,32257 and HD\,37680.

Great care should be taken when comparing results from solely
optical observations, 
as is common for extragalactic WC stars (e.g. Smartt et al. 2001),
with those additionally including UV and far-UV observations. 
In the case of HD\,37026, 
we adopt a mean oxygen abundance from the above list of diagnostics, i.e. 9\%
by mass.
The most significant model discrepancy is the predicted strength of the
O\,{\sc vi} lines -- notably $\lambda\lambda$1032-38 (for those stars with 
far-UV
datasets), $\lambda\lambda$3811-34 and $\lambda$5290. We have considered 
the inclusion
of X-rays to maintain the population of high oxygen ions 
in the wind, but without much success, especially for the optical
O\,{\sc vi} lines. We are 
only able to maintain such high ions
if the wind density were reduced,  as for Sand~2 (WO, Crowther et al. 2000), but
at the expense of the goodness of fit for other diagnostics.


Whilst the majority of spectral features are from He/C/O/Fe, there are
signatures of several other elements in the spectra of WC4 stars.
These are most readily apparent from {\it HST} and {\it FUSE} 
observations. 
Prominent P Cygni profiles of Si\,{\sc iv} $\lambda\lambda$1393-1402,
P\,{\sc v} $\lambda\lambda$1118-28 and 
S\,{\sc iv} $\lambda\lambda$1062-73 features are present, and successfully
reproduced with abundances scaled to 0.4$Z_{\odot}$.

\begin{figure}[ht!]
\vspace{7.5cm}
\includegraphics{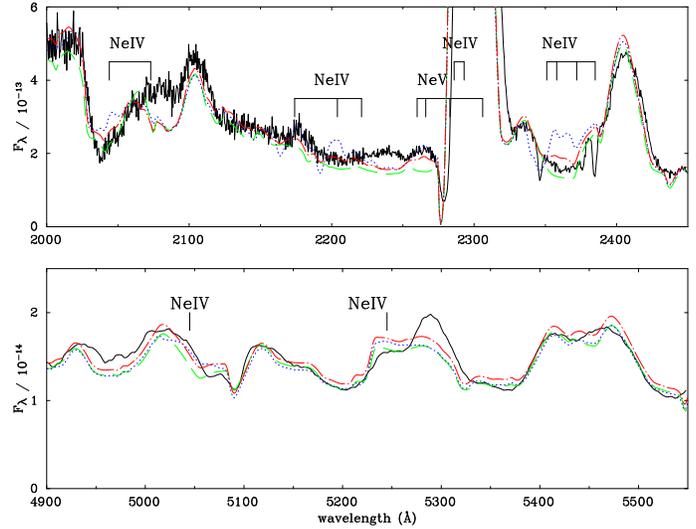}
\caption{Comparison between de-reddened observations of HD\,37026 (solid, black)
and synthetic spectra for a variety of neon abundances -- 2\% by mass
(dotted, blue), 0.4\% (dashed, red), and 0\% (dot-dashed, green).
Ordinate units are erg\,cm$^{-2}$\,s$^{-1}$\,\AA$^{-1}$.}
\label{br43-neon}
\end{figure}

Unfortunately, spectral features of neon (generally Ne\,{\sc iii-v}) in WC4 stars
are rather weak, even for large adopted abundances. This is illustrated in
Fig.~\ref{br43-neon} where UV and optical observations of HD\,37026 are compared
with models accounting for a varying amount of neon -- 2\% (dotted), 0.4\%
(dashed),  and 0\% (dot-dash). The strongest accessible (unblended) UV features
lie in the vicinity of C\,{\sc iii} $\lambda$2297 -- Ne\,{\sc iv} 
$\lambda\lambda$2043-73, $\lambda\lambda$2204-21,
$\lambda$2175, $\lambda\lambda$2351-63, Ne\,{\sc v} $\lambda$2260. These generally
favour a neon abundance of $\leq$0.4\%, which still represents a
potential overabundance by a factor of 
$\leq$5. Optical diagnostics are even harder
to locate. Ne\,{\sc iv} $\lambda$5045, $\lambda$5245 are amongst the strongest
features, but cannot readily be used to determine neon abundances without
a superior match to the observed spectrum from other species. 
We are therefore unable to verify evolutionary predictions for the 
Ne abundance in WC stars beyond those discussed by Dessart et al. (2000) 
from mid-IR fine
structure lines. 


\begin{table*}
\begin{center}
\caption{Stellar parameters for the program LMC WC4 
stars, plus Sand 2 (WO) from Crowther et al. (2000) analysed in the same
 manner. 
Note that all models assume a filling factor, $f$, of 0.1, 
whilst a two-component
velocity law was used in  all cases with $\beta_{1}=1, 
\beta_{2}=20, v_{\rm ext}=(v_{\infty} - 300)$ 
km s$^{-1}$.}
\label{table2}
\begin{tabular}{l
c@{\hspace{2.5mm}}
c@{\hspace{2.5mm}}
c@{\hspace{2.5mm}}
c@{\hspace{2.5mm}}
c@{\hspace{2.5mm}}
c@{\hspace{2.5mm}}
c@{\hspace{2.5mm}}
c@{\hspace{2.5mm}}
c@{\hspace{2.5mm}}
c@{\hspace{2.5mm}}
c@{\hspace{2.5mm}}
c@{\hspace{2.5mm}}
c@{\hspace{2.5mm}}
c@{\hspace{2.5mm}}
c@{\hspace{2.5mm}}
c@{\hspace{2.5mm}}
c}\\
\hline
HD&$T_{*}$& $R_{*}$ & $\log  L_{\rm ph}$ & T$_{2/3}$&R$_{2/3}$&
$\log \dot{M}$ & $v_{\infty}$& $\log Q_{0}$ & $\log Q_{1}$ & $\log Q_{2}$ &
C/He&O/He & $\beta_{\rm He}$ & $\beta_{\rm C}$
&$\beta_{\rm O}$ \\
    & kK    & $R_{\odot}$ & $L_{\odot}$ & kK & $R_{\odot}$&
$M_{\odot}$\,yr$^{-1}$ & km\,s$^{-1}$  &s$^{-1}$&s$^{-1}$&s$^{-1}$  &    &    & \% &\% & \%\\     
\hline
32125 & 90 & 2.2 & 5.44 & 74 & 3.2 & $-$4.8 & 2500 & 49.32&48.91 &39.03 & 0.13 & 0.04 &65&25&10\\
32257 & 85 & 2.3 & 5.42 & 71 & 3.4 & $-$4.9 & 2300 &49.29 &48.81 &36.25 & 0.35 & 0.04 &45&47& 8\\
32402 & 90 & 2.9 & 5.70 & 67 & 5.2 & $-$4.5 & 3000 &49.55 &49.10 &36.11 & 0.14 & 0.02 &66&28& 5\\
37026 & 90 & 2.8 & 5.65 & 68 & 4.8 & $-$4.5 & 2900 &49.51 &48.98 &38.80 & 0.32 & 0.05 &46&44& 9\\
37680 & 85 & 3.2 & 5.68 & 62 & 6.0 & $-$4.4 & 3200 &49.53 &49.04 &38.36 & 0.10 & 0.01 &74&22& 3\\ 
269888& 85 & 2.4 & 5.44 & 72 & 3.4 & $-$4.8 & 2600 &49.31 &48.86 &36.27 & 0.32 & 0.06 &45&43&11\\
\\
Sand 2&150&  0.65&5.28  &110 & 1.2  & $-$4.9 & 4100 &49.10 &48.85 &40.31 & 0.70 & 0.15 & 27 & 56 & 16\\
\hline
\end{tabular}
\end{center}
\end{table*}

\begin{figure} 
\vspace{13cm}
\includegraphics{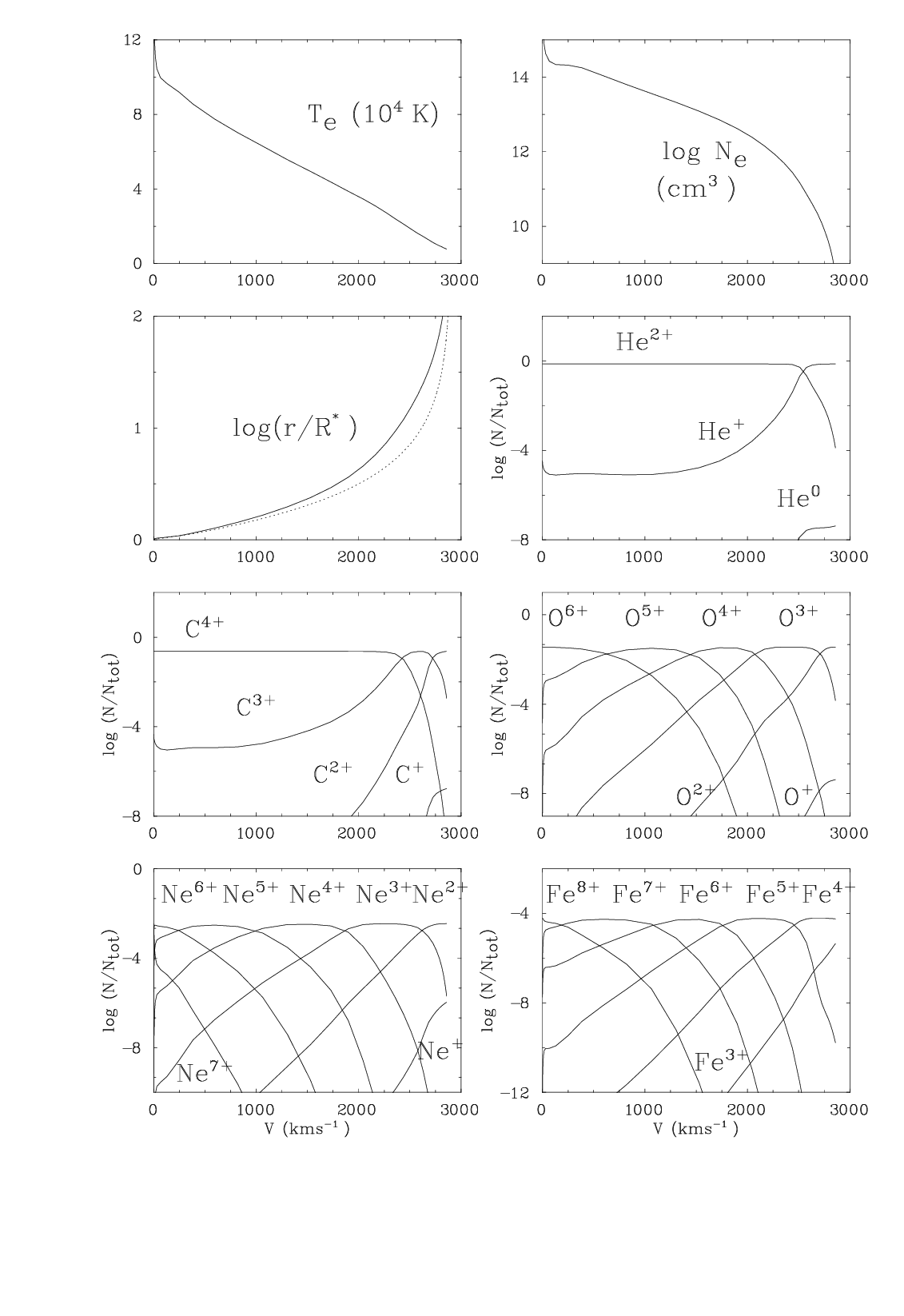}
\caption{Atmospheric structure of our final HD\,37026 model, indicating
the variation of temperature (10$^{4}$K), density (cm$^{-3}$) and radius
($R_{\ast}$) with velocity (km\,s$^{-1}$), plus the ionization balance
(in $N/N_{\rm tot}$) for helium, carbon, oxygen, neon and iron. A standard
$\beta$=1 velocity law is also included on the radius plot (dotted line).}
\label{wc4}
\end{figure}

The temperature, density, radius and ionization balance of He, C, O, Ne and Fe
for our final HD\,37026 model versus wind velocity are presented in Fig.~\ref{wc4}.
This illustrates the complex ionization structure for heavy elements with
respect to helium and carbon.

\begin{figure}[ht!] 
\vspace{9cm}
\includegraphics{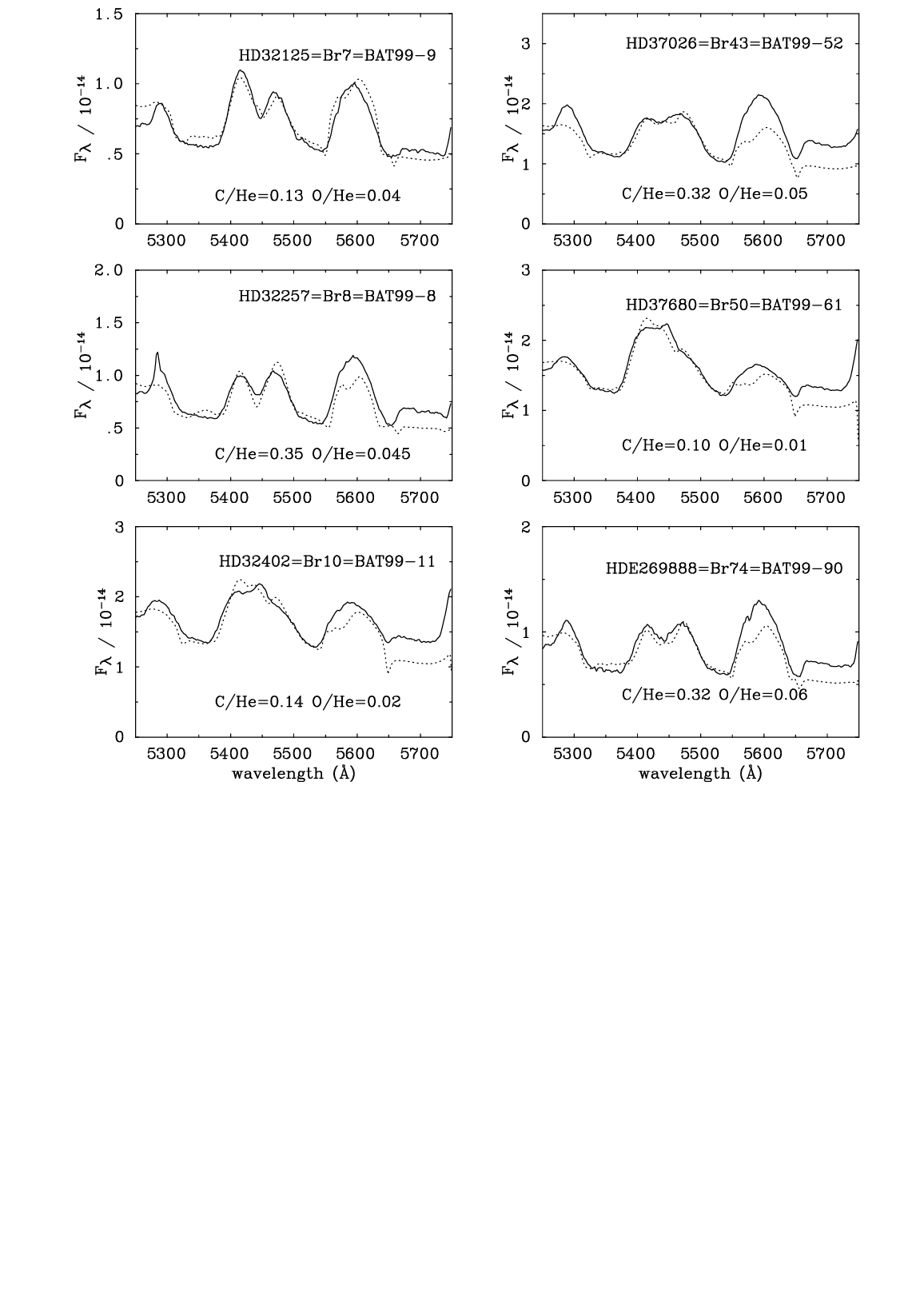}
\caption{Comparison between the de-reddened 
spectroscopy of the targets (solid) in the He\,{\sc ii} 5411, C\,{\sc iv}
5471 and O\,{\sc iii-v} 5590 region, and final theoretical predictions 
(dotted), demonstrating the excellent agreement reached for C/He 
determinations. Ordinate units are erg\,cm$^{-2}$\,s$^{-1}$\,\AA$^{-1}$.}
\label{5411_5471}
\end{figure}

\subsection{Summary of results and comparison with Gr\"{a}fener et al.}

Deduced stellar properties for each star are presented in Table~\ref{table2},
including ionizing fluxes below the H\,{\sc i} ($Q_{0}$), He\,{\sc i}
($Q_{1}$), and He\,{\sc ii} ($Q_{2}$) edges. We also present 
final synthetic spectra with each observation for the diagnostic 
He\,{\sc ii} 5412 and C\,{\sc iv} 5471 profiles in Fig.~\ref{5411_5471}. 

The present sample of stars has previously been studied by 
Gr\"{a}fener et al. (1998), whilst Hamann \& Koesterke (1998)
have considered clumped models for HD\,37026. We present a 
comparison of the derived parameters from these various studies
in Table~\ref{table3}. Overall, we obtain similar
stellar temperatures to Gr\"{a}fener et al., but substantially higher 
stellar luminosities (0.1--0.4 dex), lower wind densities 
(0.2$\pm$0.1 dex, {\it after} correcting for the clumped wind) 
and higher wind velocities (up to $\sim$15\%). 
We concur with Gr\"{a}fener et al., in that LMC WC4 stars form a 
relatively homogeneous sample, with high luminosities, temperatures
and comparable
wind densities. However, our improved observational and theoretical
approach has led to substantially revised stellar parameters, notably relating
to chemical abundances. Our improved optical spectroscopy permits
a much finer determination of C/He ratios, e.g. for HD\,37680 we obtain
C/He=0.10 by number versus C/He=0.27 according to Gr\"{a}fener et al. --
see Fig.~\ref{5411_5471}.

Differences between the two methods are as follows:
\begin{enumerate}
\item Observationally, Gr\"{a}fener et al. utilised identical
{\it HST}/FOS UV spectroscopy, plus 10--15\AA\ resolution optical datasets
from Torres \& Massey (1987) covering $\lambda\lambda$3375--7375,
with the exception of HD\,32402 for which they had higher 
quality ESO spectra. In all cases their observations 
omitted the useful far-red MSSSO and far-UV {\it FUSE} diagnostics.

\item Theoretically, similar model assumptions were made by 
Gr\"{a}fener et al., except that relatively simple model
atoms of helium, carbon and oxygen were used (e.g.
O\,{\sc iii} was omitted).  We deduce substantially 
lower O/He ratios ($<$0.1) than Gr\"{a}fener et al. 
(0.05--0.25),  plus a lower range of C/He ratios due to our 
improved  optical spectroscopy (0.1--0.35 versus 0.3--0.55) and our 
strict adherence to He\,{\sc ii} 5412/C\,{\sc iv} 5471 as abundance 
diagnostics,
plus the allowance for clumping. It is possible that erroneous C/He 
abundances may be derived if the He\,{\sc ii}
5412 electron scattering wing is unrealistically strong.

\item Line blanketing was neglected by Gr\"{a}fener et al., as was the
contribution to the opacity by iron and other heavy elements. In contrast,
we account for Ne, Si, P, S, Ar and Fe, and also
consider the (minor) influence of other heavy elements. 
The neglect of blanketing, and use of restricted model atoms
yields erroneous rectified synthetic spectra both in the UV (due
%
to the iron forest) and  the optical (due to many broad 
overlapping features), see Hillier \& Miller (1998; 1999).

\item 
Iron line blanketing is principally responsible for the increased stellar 
luminosities obtained  here relative to  Gr\"{a}fener et al. We obtain 
bolometric corrections in 
the range $-4.8 \le$ B.C. $\le -4.5$ mag, up to 0.5 mag greater than 
WC6--8 stars, but 1~mag lower than Sand~2 (Crowther et al. 2000).
We also derive somewhat higher interstellar reddenings to the program
stars than Gr\"{a}fener et al. via our spectral synthesis, typically by
$A_{\rm V}\sim$0.2 mag as presented
in Table~\ref{dered}, owing to the superior technique followed. 
\end{enumerate} 



\begin{table}
\begin{center}
\caption{Quantitative comparison between fundamental stellar parameters
newly derived here for LMC and (selected)
Galactic WC stars (referred to as C02) -- using line blanketed, 
clumped models of Hillier \& Miller 1998) -- and by Gr\"{a}fener et al. (1998,
G98) and Koesterke \& Hamann (1995, KH95) using non-blanketed homogeneous 
models, and Hamann \& Koesterke (1998 HK98) using non-blanketed clumped
models for HD\,37026. 
Catalogue number for LMC (BAT: Breysacher et al. 1999) and Galactic
(WR: van der Hucht 2001) stars are indicated in parenthesis.}
\label{table3}
\begin{tabular}{l@{\hspace{0mm}}r@{\hspace{1mm}}l@{\hspace{1mm}}c
@{\hspace{1mm}}c@{\hspace{1mm}}c@{\hspace{1mm}}c@{\hspace{1mm}}c
@{\hspace{1mm}}r}
\hline
HD &$T_{\star}$ & $\log L_{\rm ph}$
&$\log(\frac{\dot{M}}{\sqrt{f}}$) 
&$v_{\infty}$&C/He & O/He &D&Ref\\
&      kK    &  $L_{\odot}$ & 
$M_{\odot}$\,yr$^{-1}$  & km\,s$^{-1}$  &&&kpc& \\     
\hline
\multicolumn{9}{c}{LMC} \\
32125 & 90  & 5.44  &  $-$4.3 & 2500  &  0.18 & 0.04 &50&C02\\
 (9)   & 95  & 5.29  &  $-$4.2 & 2300  &  0.32 & 0.12 &50&G98\\
32257 & 85  & 5.42  &  $-$4.4 & 2300  &  0.35 & 0.05 &50&C02\\
  (8)  & 104 & 5.13  &  $-$4.1 & 2300  &  0.43 & 0.24 &50& G98\\
32402 & 85  & 5.70  &  $-$4.0 & 3000  &  0.14 & 0.02 &50&C02\\
  (11) & 92  & 5.62  &  $-$3.8 & 2800  &  0.55 & 0.16 &50&G98\\
37026 & 90  & 5.65  &  $-$4.0 & 2900  &  0.32 & 0.05 &50&C02\\     
   (52)& 103 & 5.26  &  $-$3.9 & 2600  &  0.43 & 0.24 &50&G98\\
      & 141 & 5.6   &  $-$3.7 & 2800  &  0.43 & 0.24 &50&HK98\\
37680 & 85  & 5.68  &  $-$3.9 & 3200  &  0.10 & 0.01 &50&C02\\
  (61)& 94  & 5.55  &  $-$3.7 & 2800  &  0.27 & 0.05 &50&G98\\
269888& 85  & 5.44  &  $-$4.3 & 2600  &  0.32 & 0.06 &50& C02\\   
  (90)& 104 & 5.13  &  $-$4.1 & 2300  &  0.43 & 0.24 &50& G98\\
\\
\multicolumn{9}{c}{Galactic}\\
76536 & 80  & 5.38  &  $-$4.2 & 2055  &  0.25 & 0.03 &2.0&C02\\
   (14)& 86  & 4.9   &  $-$4.2 & 1800  &  0.14 &  --  &2.0&KH95 \\
213049& 80  & 5.02  &  $-$4.5 & 2280  &  0.30 & 0.04 &2.7&C02\\
 (154)& 71  & 4.9   &  $-$4.2 & 2050  &  0.22 &  --  &3.5&KH95 \\
%
\hline
\end{tabular}
\end{center}
\end{table}

\begin{table}
\caption{Interstellar reddenings, absolute magnitudes,
and H\,{\sc i} column densities (from fits to Ly$\alpha$) to the program
LMC WC4 stars, adopting a mean Galactic foreground of $E_{\rm B-V}$ = 0.07
from Schlegel et al. (1998). Narrow-band photometry is taken from 
Torres-Dodgen \& Massey (1988), whilst we include comparisons 
with previous reddening determinations by Gr\"{a}fener et al. (1998)
and Morris et al. (1991).}                        
\label{dered}
\begin{tabular}{l@{\hspace{2mm}}r@{\hspace{2mm}}r@{\hspace{2mm}}r
@{\hspace{2mm}}r@{\hspace{-2mm}}r@{\hspace{2mm}}r}
\hline
HD& $v$ & \multicolumn{3}{c}{$E_{\rm B-V}({\rm Gal+LMC})$} & $\log 
N$(H\,{\sc i}) & $M_{v}$\\
  & mag & This work & Gr\"{a}fener & Morris & cm$^{-2}$ & mag \\
  &     &           & et al.   & et al. &            &    \\
\hline
32125  &  15.02  & 0.07+0.09 &0.03+0.08 & 0.01 & 21.5 & $-$4.02\\ 
32257  &  14.89  & 0.07+0.08 &0.03+0.06 & 0.08 & 21.0 & $-$4.12\\ 
32402  &  13.89  & 0.07+0.04 &0.03+0.05 &      & 21.0 & $-$5.00\\ 
37026  &  14.04  & 0.07+0.04 &0.03+0.01 & 0.08 & 21.0 & $-$4.85 \\ 
37680  &  14.03  & 0.07+0.06 &0.03+0.05 & 0.05 & 21.3 & $-$4.92\\ 
269888 &  15.41  & 0.07+0.27 &0.03+0.28 &      & 22.0 & $-$4.19\\ 
\hline
\end{tabular}
\end{table}

Recently, Gr\"{a}fener et al. (2002) presented a line blanketed study
of HD\,165763 (WC5) where many of their previous deficiencies 
are rectified,
i.e. enhanced model atoms for carbon and oxygen, line blanketing from
iron group elements, clumping, and a more objective line profile fitting technique.
Indeed, test calculations carried out by Gr\"{a}fener (priv. comm.) for 
HD\,37026 show excellent consistency with the present results, 
giving confidence in results obtained with the two independent codes.

Whilst Gr\"{a}fener et al. included the opacity from other iron group elements, 
Ne $\ldots$ Ar were omitted. We find that additional opacity sources
(namely Na, Mg, Cl, Ca, Cr, Mn, Ni) play negligible spectroscopic role, other than a 
strengthening of C\,{\sc iv} $\lambda\lambda$1548-51, $\lambda\lambda$5801--12 emission, 
but do contribute to the line driving force\footnote{
These additional elements are 
omitted from our present analysis for 
consistency with our recent studies of Galactic WC stars (e.g. Dessart
et al. 2000).}. From test calculations, the 
inclusion of these additional lines does help, but 
the  effect is  still insufficient to supply the necessary driving force, 
especially in initiating the outflow at the base of the wind, in common with
Gr\"{a}fener et al. (2002). This will be discussed in a forthcoming publication.


\begin{table}
\caption{Observing log for 
Galactic WC stars, including narrow-band photometry from 
Torres-Dodgen \& Massey (1988), catalogue numbers and 
cluster/association distances
from van  der Hucht (2001). Optical spectrophotometry for HD\,213049
is taken from Torres \& Massey (1987).}
\label{table1gal}
\begin{tabular}{
l@{\hspace{1mm}}r@{\hspace{1mm}}c@{\hspace{1mm}}c@{\hspace{1mm}}
c@{\hspace{1mm}}c@{\hspace{1mm}}r@{\hspace{1mm}}r@{\hspace{1mm}}r@{\hspace{1mm}}
}
\hline
HD & WR  & Sp Type & {\it IUE} & MSSSO/ & $v$ & $d$ & $E_{\rm b-v}$ & $M_{v}$ \\
   &     &         &           & KPNO & mag & kpc & mag         & mag\\
\hline
76536  & 14  & WC7  & Sep 80 & Dec 97  & 9.42 & 2.0 & 0.61 & $-$4.6\\
213049 & 154 & WC6  & May 82 & Oct 80  &11.54 & 2.7 & 0.72 & $-$3.6\\
\hline
\end{tabular}
\end{table}

\section{Comparison with properties of Galactic WC stars}\label{wc4_evolution}

We have presented a detailed study of six LMC WC stars. How
do their properties compare with Galactic counterparts? What 
differences/similarities might be expected from evolutionary predictions? 

It is well known that WC4 stars are relatively
rare in the Milky Way, with only 5 known out of 87 WC stars 
(van der Hucht 2001), whilst they dominate the LMC carbon sequence 
with 15 known out of 23 WC stars (Breysacher et al. 1999). Indeed,
most LMC WC stars with WC5--6 spectral types are  WC4+O binaries
with the later apparent subtypes resulting from 
C\,{\sc iii} $\lambda$5696 emission due to wind 
collision (Bartzakos et al. 2001).


\subsection{New analyses of Galactic WC stars}

It is unfortunate that only a few Galactic WC stars lie at known distances.
There are no  single cluster/association WC4 stars, so we are  
unable to directly compare our LMC results with Galactic counterparts. 
Instead, comparisons must be made with WC5--8 stars, which have
been analysed in a similar manner, namely
$\gamma$ Vel (WR11, WC8+O, De Marco et al. 2000); HD\,92908 
(WR23, WC6, Smartt et al. 2001), HD\,165763 (WR111, WC5, Hillier \& Miller 
1999), and
HD\,192103 (WR135, WC8, Dessart et al. 2000). To this short list we have 
added HD\,156385 (WR90, WC7) for which Dessart et al. (2000) obtained a 
distance based on a calibration of (binary) WC7 cluster/association members. 

In order to produce a more substantial set of data for Galactic WC stars,
we have newly analysed two stars whose distances are known from OB 
association membership -- HD\,76536 (WR\,14, WC7) and HD\,213049 (WR\,154, 
WC6). Spectrosopic observations, cluster distances and derived reddenings
are listed in Table~\ref{table1gal}. Ultraviolet datasets generally
represent low dispersion {\it International Ultraviolet Explorer  (IUE)} 
spectroscopy, with the exception of three (merged) high dispersion  
short wavelength (SWP) observations of HD\,76536. Optical datasets
are drawn from Mt Stromlo 2.3m DBS observations, obtained during our LMC 
Dec 1997 observing campaign for HD\,76536, plus archival 
Torres \& Massey (1987) datasets for HD\,213049 obtained with the
0.9m Kitt-Peak (KPNO) intensified Reticon scanner (IRS).
We obtain somewhat higher reddening determinations 
than the recent literature (Koesterke \& Hamann 1995; Morris et al. 1993) 
for HD\,76536 and HD\,213049, namely $E_{\rm b-v}$=0.42--0.49 and 
0.63--0.68, respectively. 

Identical analysis methods were employed to those presented above.
For conciseness, we omit presenting fits to spectral lines for HD\,76536 and 
HD\,213049, although the quality is comparable with those of HD\,156385
(WC7, Dessart et al. 2000) and HD\,92809 (WC6, Smartt et al. 2001).
Note that earlier problems with reproducing the strength of 
C\,{\sc iii} $\lambda$4186 have been resolved and were caused by
a `poor' super-level assignment.

We present derived stellar parameters in Table~\ref{table3}, together 
with those obtained using non-blanketed helium-carbon models by 
Koesterke \& Hamann (1995). The use of blanketed model atmospheres
result in higher stellar luminosities as has been discussed above for
the LMC WC4 stars, whilst mass-loss rates are again reduced even after
correcting for clumping. C/He ratios are increased in this case, although
Koesterke \& Hamann (1995) neglected oxygen in their calculations.

\begin{figure} 
\vspace{8cm}
\includegraphics{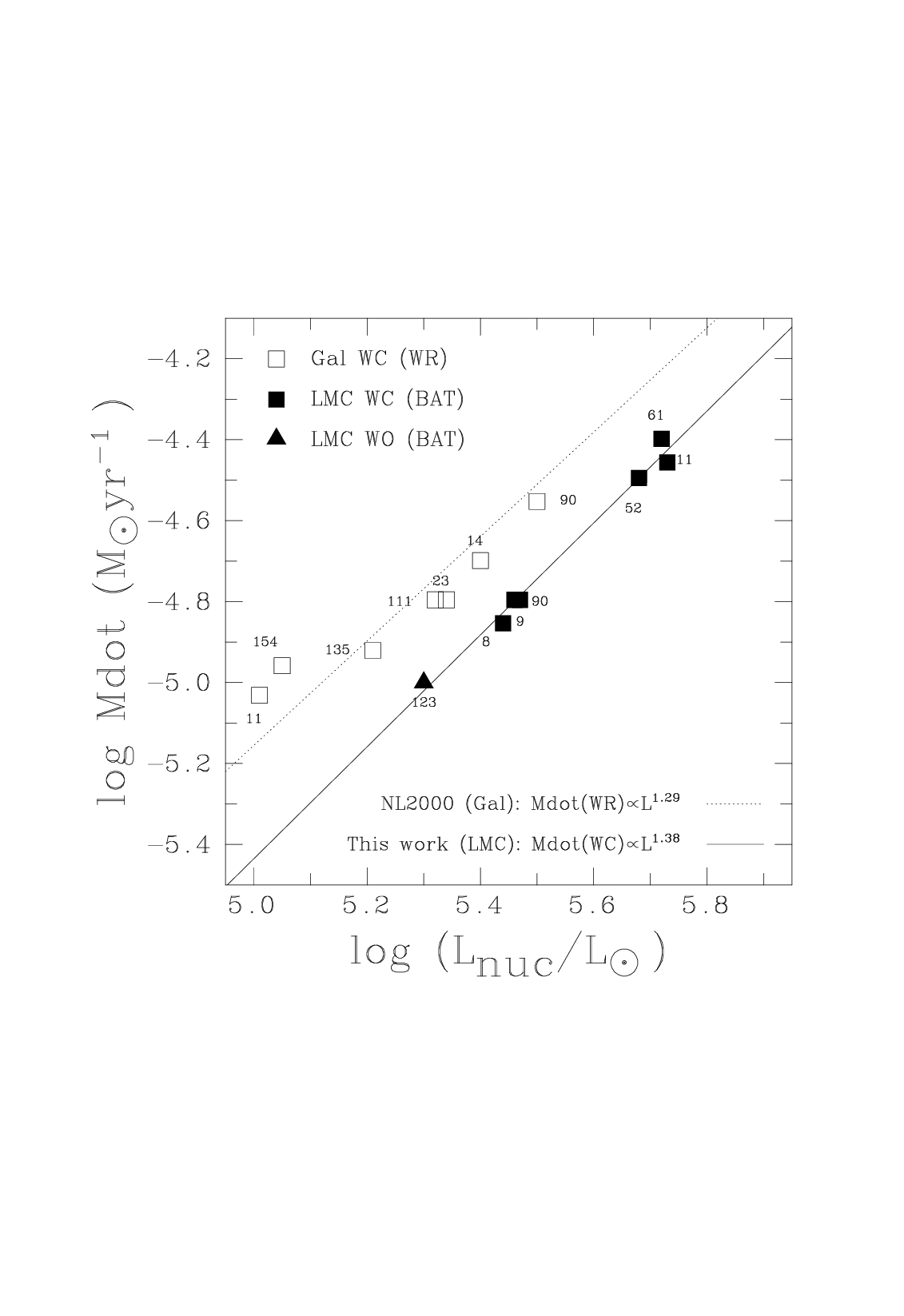}
\caption{Comparison between the mass-loss rates and luminosities
of LMC WC4 (filled squares), WO stars (filled triangle, Crowther
et al. 2000) and Galactic WC5-8 stars (open squares) at known distances. 
The lines correspond to the calibrations of Nugis \& Lamers (2000) for 
Galactic WR stars (dotted), assuming C/He=0.2, C/O=4 by
number, plus a linear fit to the LMC WC and WO results (solid).}
\label{lmc_gal}
\end{figure}

\subsection{A metallicity effect?}

In Fig.~\ref{lmc_gal} we compare the luminosities and 
mass-loss rates of
LMC WC stars with Galactic counterparts located at known distances, 
including HD\,76536 and HD\,213049 for the first time. 
Observationally, two main effects are apparent:
\begin{itemize}
\item It is striking that the {\it mean} luminosity of WC stars in 
the LMC, $\log (L/L_{\odot}$)=5.6, is a factor of two times higher than 
those of their Galactic counterparts, $\log (L/L_{\odot}$)=5.3, 
which have been studied in an analogous manner. The conventional
interpretation of this difference is 
that LMC stars are descended from, on average, 
higher initial mass stars, as might be expected from evolutionary predictions
that account for the lower mass-loss rates during the main-sequence and 
post main-sequence phases at lower metallicities. A more speculative 
alternative is that their initial 
masses span a similar range, but that they are currently faster rotators
owing to lower mass-loss rates during earlier evolutionary phases. 
Polarization evidence points to the majority of Galactic WC stars being
fairly symmetric, and hence are slow rotators by inference (Harries et 
al. 1998).
\item LMC WC stars possess lower mass-loss rates than Galactic stars of 
similar luminosity, unless clumping factors differ between the two samples. 
The semi-empirical calibrations of Nugis \& Lamers (2000) for Galactic WR
stars match the observed WC properties rather well, whilst a linear fit 
to the LMC WC and WO data reveals

\begin{equation}
\log (\dot{M}) = 1.38  \log (L/L_{\odot}) - 12.35
\end{equation}

with a similar slope to the Nugis \& Lamers (2000) generic
WR calibration (their Eqn.~22), albeit offset by $\sim$0.2 dex. 
\end{itemize}

The metallicity dependence of the mass-loss rate for O stars
($\dot{M} \propto Z^{0.5-0.7}$) is well established observationally
(Puls et al. 1996) and theoretically (Kudritzki \& Puls 2000;
Vink et al. 2001). 
The situation for WN stars is less clear (Crowther 2000; Hamann \& 
Koesterke 2000). However, since iron group elements provide most of
the driving for the winds of WN stars, they should exhibit a similar
dependence on metallicity as the O-type stars.
A metallicity dependence of $\sim Z^{0.5}$ would predict that
LMC WR stars have winds that are 0.2 dex lower than equivalent Galactic stars,
as appears to be the case.

One might question whether a metallicity dependence of WC stars is
expected, since their winds are so 
dominated by carbon and oxygen. From test calculations for WC4 stars, 
we find that the contribution of lines from dominant ions 
($\sim$250 eV) at the  base of the wind ($<$ 100 km\,s$^{-1}$) is 
crucial to initiating the outflow. Carbon does not possess 
lines from such ions since the ionization potential of C$^{3+}$ (64 eV)
and C$^{4+}$ (392 eV) do not lie in the necessary range -- the same is
true of oxygen (recall Fig.~\ref{wc4}). 
Instead, ions from Ne, Ar, and Fe-group elements appear to be 
critical in initiating the outflow from early-type WC stars, which
would qualitatively explain their sensitivity to heavy metal content.

\begin{figure} 
\vspace{12cm}
\includegraphics{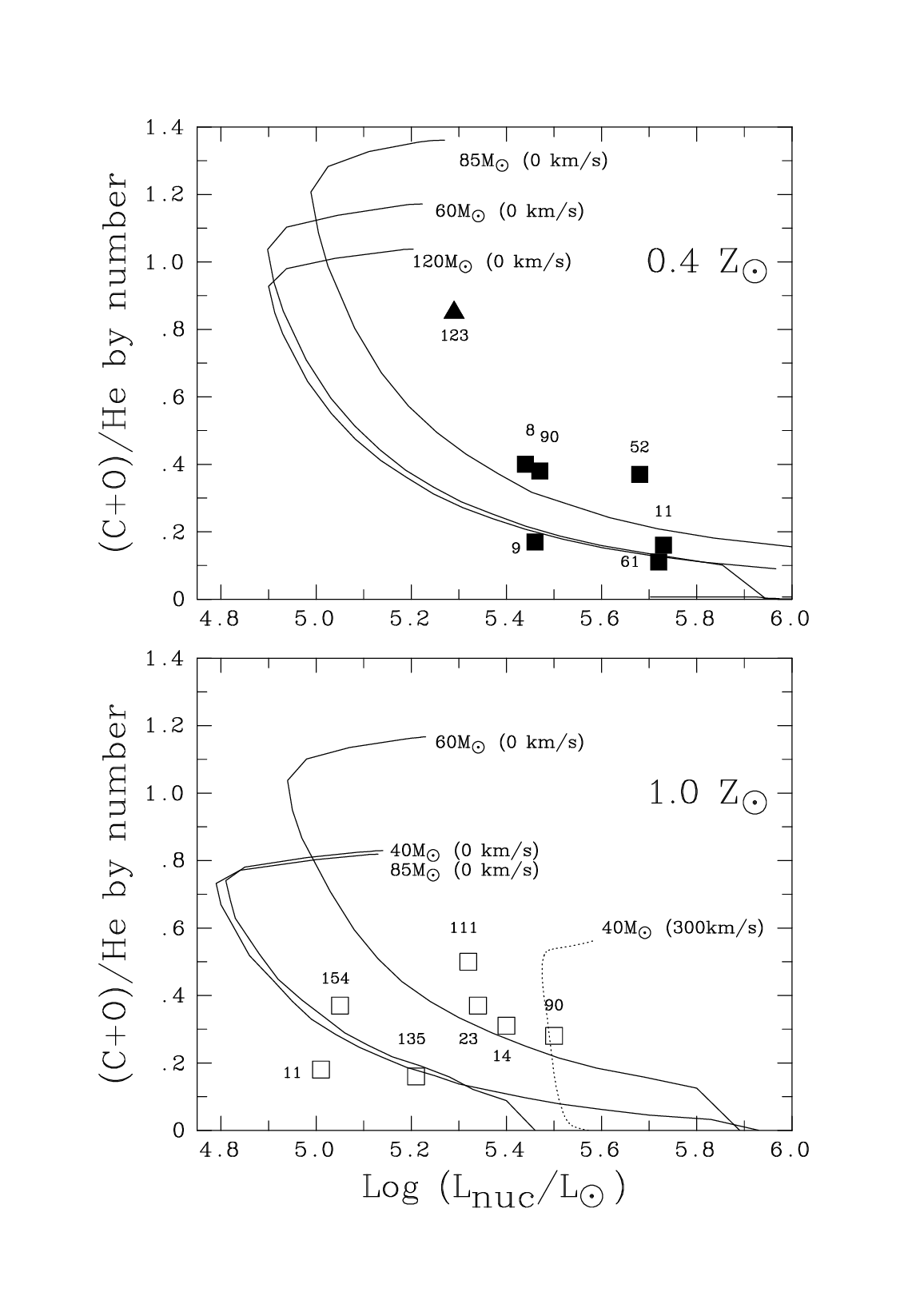}
\caption{Comparison between the (C+O)/He values of WC stars 
in the LMC and Galaxy versus stellar luminosity, together with
evolutionary predictions for non-rotating stars (Meynet et al.
1994) for the two metallicities, plus results from an initially
rotating 40$M_{\odot}$ solar metallicity model from Meynet \& Maeder (2000). 
Symbols (and BAT/WR numbers) are as for the previous figure.}
\label{co}
\end{figure}

We shall see in the next section that a metallicity dependence of 
mass-loss rates contributes to an explanation for the
subtype distribution amongst WC stars. In addition, since weaker
WR winds are more transparent to harder ionizing photons (Schmutz et 
al. 1992; Crowther 1999) one would expect a greater contribution
of WR stars to the hard ionization of young starburst regions in
more metal deficient galaxies. The inclusion of line blanketed
metallicity dependent mass-loss rates for WR stars in evolutionary
synthesis calculations is discussed by Smith
et al. (2002).

The Galactic and LMC WC populations are further compared in
Fig.~\ref{co}, where we present their (C+O)/He values versus luminosity,
together with evolutionary predictions for non-rotating stars 
(solid, Meynet et al. 1994) for the two metallicities, plus results 
for an initially rapidly rotating 40$M_{\odot}$ model at 1.0$Z_{\odot}$
(dotted, Meynet \& Maeder 2000). It is apparent that these stars 
reveal similar surface enrichments, despite their 
differences in spectral types (Galactic: WC5--8; LMC: WC4) and stellar 
luminosities. The shift to (observed) higher luminosity is reasonably 
well reproduced in the non-rotating evolutionary models, although 
the shift in the (sole) rotating track is dramatic. Therefore, we must await
results for a wider initial range of masses and rotational velocities 
before we can have confidence in the validity of these predictions.

\subsection{Differences in WC spectral types between Galactic and LMC stars?}

While we have established that WC stars in the LMC are typically more luminous,
and possess lower mass-loss rates than Galactic counterparts, what is the origin
of their systematically earlier spectral types,?
This has been assumed to be due either
to temperature differences and/or an abundance effect.

Smith \& Maeder (1991) adopted the hypothesis that the 
difference between the WC subtypes was primarily governed 
by a variable surface  abundance ratio, (C+O)/He. This
was based on recombination line studies of a 
sample of Galactic WC4--9 stars. However, only a single
WC4 star was included in their sample, whilst recombination
studies of late-type WC stars are fraught with difficulties
(Hillier 1989). The present results reveal  further that
the expected (C+O)/He ratios for WC4 stars (0.7--1.0) 
are not supported here (0.1--0.4), with all Galactic and LMC
WC stars poorer in carbon and oxygen than standard (non-rotating) 
evolutionary models would suggest.

With regard to oxygen abundances, evolutionary predictions for non-rotating 
massive stars indicate O/C=0.1--0.22 by number at the stage in which 
0.1$\le$ C/He $\le$0.3, in good agreement with 
O/C=0.1--0.25 determined 
here. 
Predictions from  evolutionary models in which rotation is 
considered (albeit for solar abundances  only) 
indicate a fairly similar range (O/C=0.2--0.3 by number).

\begin{figure*} 
\vspace{22cm}
\includegraphics{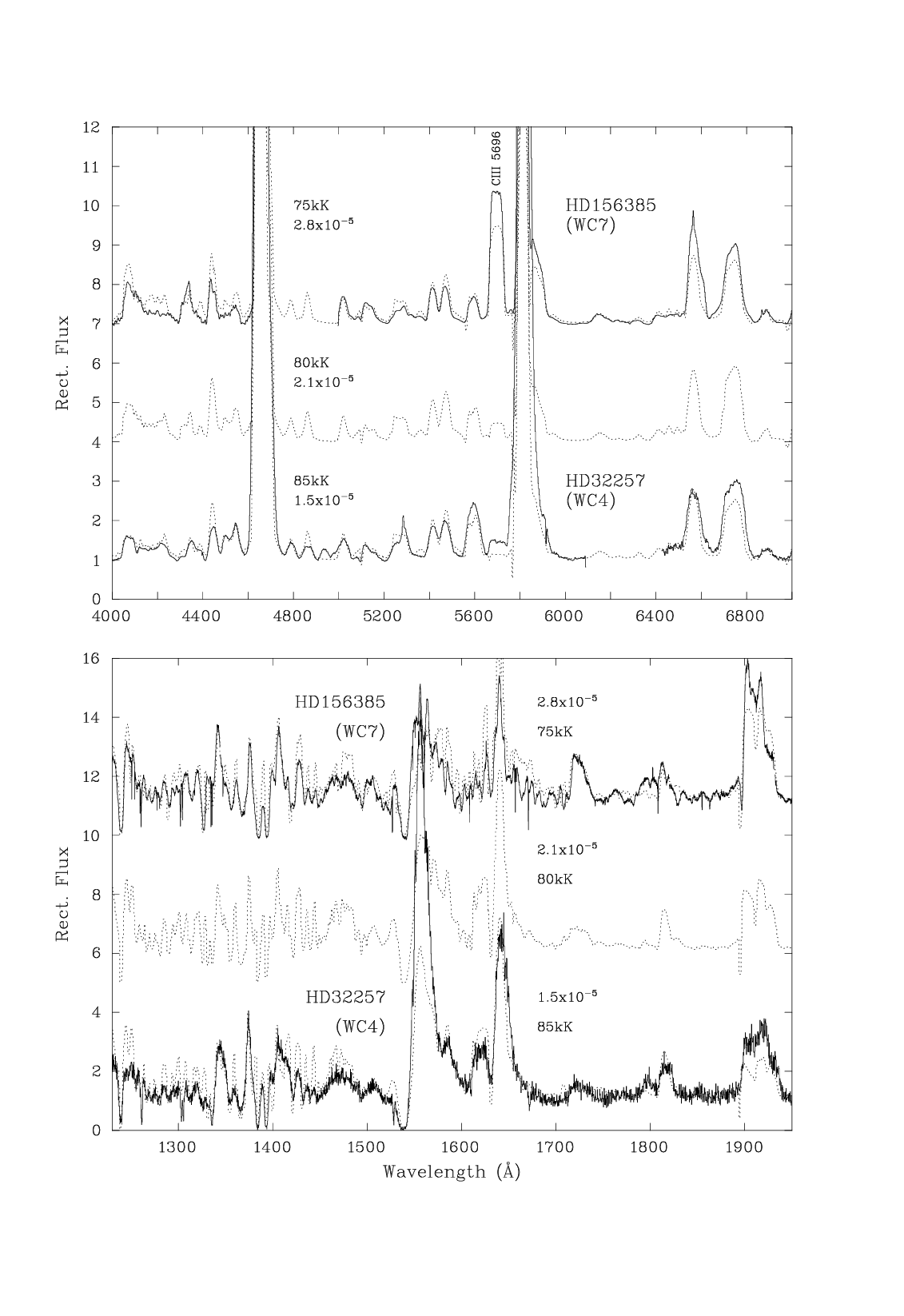}
\caption{Rectified optical (upper panel) and UV (lower panel)
spectroscopy of HD156385=WR90 (WC7, Galactic) and 
HD\,32257=Br\,8 (WC4, LMC), together with synthetic spectra from a
series of three models, with log ($L/L_{\odot}$)=5.49, 
$v_{\infty}$=2300 km\,s$^{-1}$, C/He=0.35 and O/He=0.06, except that 
(bottom to top) the temperature:mass-loss rate is 75kK:2.8$\times 10^{-5}$
$M_{\odot}$yr$^{-1}$, 
80kK:2.1$\times 10^{-5}$
$M_{\odot}$yr$^{-1}$, and 85kK:1.5$\times 10^{-5}$$M_{\odot}$yr$^{-1}$, 
 respectively. 
Remarkably, these small changes in stellar parameters closely match
the optical and UV spectra of these WC4 and WC7 stars, with C\,{\sc iii}
$\lambda$5696 most sensitive to small changes in physical parameter
(compare $\lambda$5696 with $\lambda$6740 in the upper panel).}
\label{wc4_wc7}
\end{figure*}

Crowther (1999) proposed that the variety of early-type WC 
subtypes results  principally from 
differing wind densities, rather than surface abundances
or ionization. This seems to 
be supported in Fig.~\ref{lmc_gal}. We have further illustrated this effect
in Fig.~\ref{wc4_wc7} where we present UV and optical
observations of HD\,156385 
(WC7, Galactic) and HD\,32257 (WC4, LMC). 
Clearly these distinct spectral types do not differ dramatically 
in their optical spectrum,  other than the strength of the 
C\,{\sc iii} classification diagnostic at $\lambda$5696. Hillier
(1989) first emphasised the sensitivity of this line to stellar
parameters, owing to the alternative decay mechanism from the upper level
to 574\AA\ with a branching ratio of $\sim$150. Consequently, 
$\lambda$5696 will only be strong when $\lambda$574 has a large optical
depth. Note that other
C\,{\sc iii} lines, for instance the feature at $\lambda$6740, do not differ
significantly between the two stars. Conti et al. (1990) has 
previously remarked upon how
other strong C\,{\sc iii} lines ($\lambda$4650, $\lambda$9710)
do not closely follow the classification C\,{\sc iii} line. 
In addition, the He\,{\sc ii}
$\lambda$5412--C\,{\sc iv} $\lambda$5471 blend is remarkably similar 
for these
two stars, such that their C/He abundances are essentially identical.

We include synthetic spectra for HD\,32257 (from 
the present work) in Fig.~\ref{wc4_wc7}, plus two further models in which changes to
the temperature and mass-loss rate are made. The temperature is reduced 
from 85kK to 80kK, and then to 75kK, whilst the mass-loss 
rate is increased from 1.5 $\times 10^{-5} M_{\odot}$ yr$^{-1}$, to 
2.1 $\times 10^{-5} M_{\odot}$ yr$^{-1}$, and then to
2.8 $\times 10^{-5} M_{\odot}$ yr$^{-1}$, in order to closely match the 
parameters determined by Dessart et al. (2000) for HD\,156385. 
From the figure, it is apparent that this small shift in parameters is 
sufficient for the spectral type to shift from WC4 to WC7, whilst
other C\,{\sc iii} lines remain  remarkably stable. Stellar temperature
certainly plays a role - fixing the mass-loss rate and increasing the
luminosity at constant radius also reduces the strength of $\lambda$5696. 
In the UV, the effect of varying wind density is also evident. For example, the
higher mass-loss rate of 
HD\,156385 dramatically weakens C\,{\sc iv} $\lambda\lambda$1548--51 through 
iron-group blanketing. 

From Fig.~\ref{wc4_wc7} one might presume that mass-loss plays
a principal role in determining spectral types for all WC stars,
via the effect on C\,{\sc iii} $\lambda$5696, However 
this is not always the case.
It has been established
for several decades that 
early-type WC stars possess broader lines than late-type WC stars. 
Whatever its physical origin, 
this also has the effect of weakening the classification 
C\,{\sc iii} $\lambda$5696 line. From recent, 
mostly unpublished calculations, it is apparent that WC8 and 
especially WC9 stars do have significantly lower stellar temperatures
than  other WC subtypes. Wind density does appear to 
play the principal role amongst WC4--7 stars, 
because of the sensitivity of the classification 
C\,{\sc iii} $\lambda$5696 line to mass-loss rate. Therefore, 
the spectral type amongst WC4--7 stars, which we consider to 
represent `early-type' WC stars is 
principally defined by wind density, with temperature a secondary effect
and no role played by carbon abundance. WC4 and WO stars {\it definitely}
possess weaker winds than other WCE stars, and {\it probably} 
have the highest stellar temperatures. Most likely, weaker winds {\it and}
higher temperatures distinguishes WO stars from WC4 subtypes.

In comparison, the classification criteria for WN stars are much stronger
indicators of ionization than those of WC stars. The sole exception is 
that N\,{\sc iv} $\lambda$4058/N\,{\sc iii} $\lambda$4634--41 is sensitive
to nitrogen abundance at fixed temperature, as illustrated by
Crowther (2000). This ratio now forms the basis for classification of
early O stars (Walborn et al. 2002a).


\section{Wolf-Rayet Masses}

\begin{table}
\begin{center}
\caption{Determinations of masses for Galactic 
and LMC WC stars using the relationship from 
Langer (1989) for stellar (photon) luminosities ($M_{\rm L89}$). 
Correction for wind blanketing effects discussed by
Heger \& Langer (1996) imply slightly higher (nuclear) luminosities 
($L_{\rm nuc}$)
and  masses ($M_{\rm nuc}$). We also provide predicted final pre-SN
masses ($M_{\rm pre-SN}$) based on estimates of lifetimes ($\tau$) 
remaining,
using  alternative evolutionary models - Meynet et al. (1994, 0.4$Z_{\odot}$
and 1.0$Z_{\odot}$) or Meynet \& Maeder (2000, 1.0$Z_{\odot}$) as discussed
in the text, 
plus the time dependent mass-loss rates scaled according to Eqn.~1.
Stellar properties for Galactic stars are taken from Hillier \& Miller (1999),
Dessart et al. (2000), De Marco et al. (2000), Smartt et al. (2001).}
\label{table4}
\begin{tabular}{l@{\hspace{-2mm}}r@{\hspace{0.5mm}}r@{\hspace{2mm}}r@{\hspace{2mm}}
c@{\hspace{2mm}}c@{\hspace{2mm}}c@{\hspace{-5mm}}r}
\hline
HD & $\log L_{\rm ph}$ & $\log L_{\rm nuc}$ & 
$M_{\rm L89}$ & $M_{\rm nuc}$ & 
$\tau$            & $\dot{M}$ &
$M_{\rm pre-SN}               $\\ 
   &$L_{\odot}$& $L_{\odot}$ & $M_{\odot}$ & $M_{\odot}$ & 
$10^{5}$yr & 10$^{-5} M_{\odot}$\,yr$^{-1}$  & $M_{\odot}$ \\     
\hline                                          
%
32125 & 5.44 &5.46&  14.6 &14.8& 3.7        & 1.6--0.7 &  11.1  \\   
(WC4) &      &    &       &    & 0.9        & 1.6--1.2 &  13.7  \\   
32257 & 5.42 &5.44&  14.3 &14.3& 2.7        & 1.4--0.8 &  11.8  \\  
(WC4) &      &    &       &    & 0.5        & 1.4--1.2 &  14.0  \\  
32402 & 5.70 &5.73&  20.4 &21.2& 4.0        & 3.5--1.1 &  13.8  \\  
(WC4) &      &    &       &    & 1.0        & 3.5--2.4 &  18.7  \\  
37026 & 5.65 &5.68&  19.1 &19.8& 2.9        & 3.2--1.3 &  14.1  \\  
(WC4) &      &    &       &    & 0.6        & 3.2--2.5 &  18.2  \\  
37680 & 5.68 &5.72&  19.9 &20.9& 4.5        & 4.0--0.9 &  12.2 \\   
(WC4) &      &    &       &    & 1.1        & 4.0--2.5 &  17.5 \\   
269888& 5.45 &5.47&  14.8 &15.0& 2.7        & 1.6--0.8 &  12.1  \\   
(WC4) &      &    &       &    & 0.5        & 1.6--1.4 &  14.4  \\   
\\
68273 & 5.00 &5.01&   8.9 & 8.7& 3.9        & 0.9--0.4 &  6.5  \\
(WC8) &      &    &       &    & 0.9        & 0.9--0.7 &  8.1   \\   
76536 & 5.38 &5.40& 13.6  &13.7& 3.0        & 2.0--0.8 &  9.7   \\   
(WC7) &      &    &       &    & 0.7        & 2.0--1.6 & 12.4   \\   
92809 & 5.32 &5.34& 12.7  &12.7& 2.5        & 1.6--0.8 &  9.8   \\
(WC6) &      &    &       &    & 0.5        & 1.6--1.4 & 12.0   \\   
156385& 5.48 &5.50& 15.4  &15.5& 3.0        & 2.5--1.0 & 10.6   \\
(WC7) &      &    &       &    & 0.7        & 2.5--2.0 & 13.9   \\   
165763& 5.30 &5.32& 12.4  &12.5& 1.8        & 1.6--0.9 & 10.1   \\
(WC5)  &      &    &       &    & 0.1        & 1.6--1.6 & 12.3   \\   
192103& 5.20 &5.21& 11.0  &10.9& 4.1        & 1.2--0.5 &  7.9   \\
(WC8) &      &    &       &    & 1.0        & 1.2--0.9 & 10.0   \\   
213049& 5.02 &5.05& 9.1   & 9.1& 2.5        & 1.2--0.6 &  7.0   \\   
(WC6) &      &    &       &    & 0.5        & 1.2--1.0 &  8.5   \\   
\hline
\end{tabular}
\end{center}
\end{table}

\subsection{Present and final masses of WC stars}

Present LMC WC4 masses can be calculated based on their stellar 
luminosities, according to Schaerer \& Maeder (1992) or Langer (1989),
revealing values in the range 12--18$M_{\odot}$. 
However, these calculations neglect 
the fact that measured (photon) luminosities in WR stars 
will be less than that generated by nuclear processes in the 
deep interior, the so-called `wind darkening' effect. Taking wind 
darkening into account results in a slightly larger mass 
(Heger \& Langer 1996). These calculations 
are presented in Table~\ref{table4}, together with masses derived from
Langer (1989). Note that the differences are
much smaller than those reported previously due to the reduced
mass-loss rates for WR stars once clumping is taken into account.
We have confidence in our derived masses due to the close agreement
between the mass of the WC8 component in $\gamma$ Vel obtained
here from a previous spectral analysis ($\sim$9$M_{\odot}$) with that
obtained from an orbital study of this binary system
(9.5$M_{\odot}$, De Marco et al. 2000). 

Further, in order to estimate final masses we make the following 
assumptions:
\begin{itemize}
\item Lifetimes remaining 
prior to a supernova explosion can be adopted from recent
evolutionary models. For example, HD\,32125 (for which 
C/He=0.18 by number) is predicted to end its life in 3.7$
\times 10^{5}$ yr according to the 0.4$Z_{\odot}$ tracks from 
Meynet et al. (1994)  irrespective of its initial mass (in the range 
60--120$M_{\odot}$). Alternatively, Meynet \& Maeder (2000) have 
presented evolutionary tracks for rotating (solar metallicity) massive 
stars -- following their predictions at 40$M_{\odot}$ a much lower 
remaining age is obtained -- 0.9$\times 10^{5}$ yr. Since the dependence
of metallicity and initial mass remains to be established for rotating models
we consider both possibilities in Table~\ref{table4}. 
\item We adopt a mass (luminosity) dependent mass-loss rate for the remainder
of the star's life, scaled according to Eqn.~1 and using the Langer (1989) 
mass-luminosity relationship. For the above example with standard
evolutionary tracks, the mass-loss rate
is reduced from 1.6$\times 10^{-5}$ to 0.7$\times 10^{-5}$ 
$M_{\odot}$ yr$^{-1}$ over the next 3.7$\times 10^{5}$ yr, ending
its life with a mass of 11.1$M_{\odot}$.
\end{itemize}
 
Consequently, we consider final, pre-SN masses of 11--19$M_{\odot}$ for 
the entire
LMC sample. Given the systematically less luminous Galactic stars, we
determine correspondingly lower final masses of 7--14$M_{\odot}$ for these
stars. 

In addition, we are able to estimate the total amount of carbon released
by the star throughout its WC lifetime which we have calculated as follows. 
For each epoch we have determined the mass fraction of carbon,
$X_{\rm C}$, via quadratic or linear fits to predictions of $X_{\rm C}$
versus remaining lifetime from Meynet et al. (1994) or Meynet \& 
Maeder (2000). The amount of carbon liberated is then simply the product 
of the age interval, age dependent mass-loss rate (see above) and carbon 
mass fraction. In the above example, HD\,32125 is predicted to lose a total
of 2.2$M_{\odot}$ of carbon during its WC
phase using figures from
Meynet et al. (1994) or 0.6$M_{\odot}$ from Meynet \& Maeder (2000).
Our six WC4 targets are predicted to provide the LMC interstellar medium 
with 5--20 $M_{\odot}$ of carbon over their WC lifetimes.

\subsection{What masses for Type Ic SN?}  

It is possible to determine lower limits to the CO masses of
Type Ic SN from modelling their light curves and fitting expanding models 
to their early spectra. For example, 
Iwamoto et al. (1998, 2000) present fits to 
three SN Ic's, revealing CO core masses in the 
range 2.1$M_{\odot}$ (SN\,1994I), 10$M_{\odot}$ (SN\,1997ef),
13.8$M_{\odot}$ (SN\,1998bw).   Recently, Smartt et al. (2002) have
discussed potential precursors to the type Ic SN\,2002ap in M74,
which include a WC star. From an examination of pre-SN images of this
galaxy (with an LMC like metallicity), they deduced an upper limit of 
$M_{\rm B}=-$6.3$\pm$0.5 mag
for the immediate pregenitor. The brightest of our six WC stars  
has $M_{b}=-5.3$ mag, which corresponds to $M_{\rm B}=-5.8$ mag in
the Johnson system. None of these stars would have been visible on
pre-SN images, adding substance
to a possible WC precursor to SN~2002ap. 

The values presented above
represent minimum values of the final progenitor 
mass,
since an unknown fraction may have imploded to form a black hole or
other compact object.  
Nevertheless, the CO ejecta mass of SN\,1998bw fits in remarkably well 
with the predicted final masses of LMC WC stars determined here. 
Evolutionary predictions, plus model analyses, imply that He represents 
a non-negligible fraction  of the final surface abundance of very 
massive stars. This is consistent with the initial 
He-free Type-Ic classification for SN\,1998bw, since subsequent 
observations did reveal the presence of 
He\,{\sc i} 1.083$\mu$m (Patat et al. 2001).

SN\,1998bw was also remarkable 
in that it appeared to be co-incident with GRB\,980425. This Gamma Ray 
Burst was relatively sub-luminous, and it has been argued by 
Woosley et al. (1999) that such outbursts are atypical for GRBs. 
Nevertheless, SN\,1998bw/GRB\,980425 most likely corresponds to the 
collapsed core of a very massive star at the end of its WC phase. 
Recent XMM-Newton observations of GRB\,011211 by Reeves et al. (2002)
provides further support for such an origin due to enriched Mg, Si, S,
Ar, Ca and Ni in the X-ray afterglow.

The lower CO mass of SN\,1997ef is consistent with that of a 
luminous WC stars in our Galaxy, such as HD\,165763 (WR111, WC5)
for which we obtain a final mass of 10--12$M_{\odot}$. 
The exceptionally low CO mass of SN\,1994l has no analogue amongst
Galactic or LMC massive WC stars, unless most of the final
CO mass imploded as a black hole/neutron star.


       \section{Conclusions}

We have used far-UV, UV, optical and near-IR spectroscopy together
with line blanketed model atmospheres to re-evaluate the stellar
parameters of six LMC WC4 stars. We derive stellar luminosities 
which are a factor of $\sim$2 higher than previously 
established for these stars (Gr\"{a}fener et al. 1998),
and systematically higher than Galactic counterparts at known distance. 
Mass-loss rates are lower than previously determined due to
the inclusion of clumping in our models. Several lines of evidence,
in both WN and WC stars, strongly suggest that the winds of WR are
strongly clumped.

We find that the LMC WC4 stars 
possess very similar surface abundances to Galactic WC5--8 stars, 
although their wind densities are systematically lower, by $\sim$0.2 dex.
We propose that the lower heavy metal content of the LMC is responsible
for their lower mass-loss rates via their (still unproven) radiatively 
driven winds. A metallicity dependence of $\sim Z^{0.5}$ would produce this 
weak effect, comparable to that predicted for O stars
as a function of metallicity ($\dot{M} \propto Z^{0.5-0.7}$: Kudritzki \& 
Puls 2000;
Vink et al. 2001). 
This relatively minor difference quantitatively explains their 
different subtype distribution (recall Fig.\ref{wc4_wc7}). Temperature 
appears to be a secondary criteria in distinguishing WC4--WC7 stars.
Late-type WC stars, especially WC9, do appear to be rather different, with 
systematically lower stellar temperatures.  
Looking at the broader picture of the subtype  distribution in other 
Local Group galaxies, other factors clearly come into play. 

If metallicity were the sole factor which determined the
wind strengths of WC stars, those galaxies with metallicities close 
to Galactic (e.g. M33 and M31) would have a large, late-type WC population, 
which is apparently not the case (Massey \& Johnson 1998), 
unless these are all obscured in the visible 
by circumstellar dust. Recent evidence suggests that
most (or all) WC9 stars are close binaries with time-dependent
dust production from interactions between the winds of the
WC and OB companion 
(e.g. Tuthill et al. 1999). Future quantitative studies of the properties 
of these and more distant WR stars will hopefully help address these issues.
Regardless of these details, 
the identification of a heavy metallicity dependence of 
WR winds has a major impact on their relative contribution 
to the hard ionizing photons in young starburst galaxies (Smith et al. 2002).

The association between spectral type and wind density amongst LMC and
Galactic WCE stars also has relevance to the long standing debate amongst
[WC]-type central stars of Planetary Nebulae (CSPN). It has long been
recognised that [WC] stars positively avoid WC5-7 spectral types, and
either cluster around WC4/WO or WC8--10 (e.g. Fig.~2 in 
Acker et al. 1996; Crowther et al. 1998). A wind density origin 
for the differences between WC4--7 stars (rather than
temperature) can therefore explain the observed clustering as due to
reduced wind densities amongst [WC]-type CSPN relative to massive Galactic
WC stars, without the need to infer rapid evolution, for instance.
Clearly, one still would need to understand why [WC]-type CSPN
possess weaker winds than their massive cousins, but the differences
would not need to be great -- a factor of $\sim$2 would suffice.

Finally, we determine current masses of LMC and Galactic WC stars 
including the 
(minor) effect of wind darkening (Heger \& Langer 1996). Based on 
remaining lifetimes from evolutionary predictions we estimate final 
pre-SN masses of 11--19$M_{\odot}$ for LMC WC stars, and 7--14$M_{\odot}$ 
for Galactic WC stars with known distances. These values are consistent 
with WC stars being the immediate precursors to luminous Type-Ic
supernova explosions, including SN\,1998bw and SN\,1997ef.

\begin{acknowledgements}

{\it FUSE} is operated for 
NASA by the Johns Hopkins University under NASA contract NAS5-32985. 
The NASA/ESA Hubble Space Telescope is operated by AURA under the NASA
contract NAS5-26555. Financial support is acknowledged from the Royal 
Society (PAC) and the UCL Perren Fund (LD). JDH acknowledges support by
NASA through grants 4450.01-92A, NAWG-3828, plus 5460.01-93A from the 
STScI. Part of this research was carried out during a visit by JDH to 
London, which was funded via the PPARC grant PPA/V/S/2000/00500. 
We are indebted to the
members of the OPACITY project, IRON project, plus Bob Kurucz, for their 
continuing endeavours to calculate atomic data without which studies presented
above would not be possible.
We are grateful to Kenneth Brownsberger for allowing his 
Guaranteed {\it FUSE} Time to be used for the study of Wolf-Rayet stars, 
to Jason Tumlinson for the use of his IDL-based molecular 
hydrogen fitting tool, and to Goetz Gr\"{a}fener for calculating a
test model using the current version of his model atmosphere code. 
Finally, thanks to the referee, Agnes Acker, for valuable comments.

\end{acknowledgements}

\end{document}